\documentstyle[11pt,epsf]{article}
\newif\ifnoepsf 
\ifnoepsf\else\input{epsf.sty}\fi
\newcommand{\figinclude}[2]%
{\ifnoepsf\fbox{figure {\tt #2} goes here}%
\else{\epsfxsize=#1 pt \epsfbox{#2}}\fi}
\newlength{\bredde}
\def\slash#1{\settowidth{\bredde}{$#1$}\ifmmode\,\raisebox{.15ex}{/}
\hspace*{-\bredde} #1\else$\,\raisebox{.15ex}{/}\hspace*{-\bredde} #1$\fi}

\newcommand{\beq}{\begin{equation}}
\newcommand{\eeq}{\end{equation}}
\newcommand{\noi}{\vspace{12pt}\noindent}
\newcommand{\lG}{\raise.3ex\hbox{$\stackrel{\leftarrow}{G}$}}
\newcommand{\lU}{\raise.3ex\hbox{$\stackrel{\leftarrow}{U}$}}
\newcommand{\lP}{\raise.3ex\hbox{$\stackrel{\leftarrow}{{\cal P}}$}}
\newcommand{\leta}{\raise.3ex\hbox{$\stackrel{\leftarrow}{\eta}$}}
\newcommand{\lOmega}{\raise.3ex\hbox{$\stackrel{\leftarrow}{\Omega}$}}
\newcommand{\ldr}{\raise.3ex\hbox{$\stackrel{\leftarrow}{\delta^r}$}}
\newcommand{\cn}{c_n^2}
\newcommand{\cnp}{c_{n+1}^2}
\newcommand{\cnm}{c_{n-1}^2}
\def\beqn{\begin{eqnarray}}
\def\eeqn{\end{eqnarray}}
\def\rar{\rightarrow}

\def\gtwid{\raise.3ex\hbox{$>$\kern-.75em\lower1ex\hbox{$\sim$}}}
\def\ltwid{\raise.3ex\hbox{$<$\kern-.75em\lower1ex\hbox{$\sim$}}}
\def\l{\label}
\def\r{\ref}
\def\la{\lambda}
\def\al{\alpha}

\begin{document}
\topmargin -1.4cm
\oddsidemargin -0.8cm
\evensidemargin -0.8cm
\title{\Large{{\bf Multicritical Microscopic Spectral Correlators \\of
Hermitian and Complex Matrices}}}

\vspace{0.5cm}

\author{
{\sc G. Akemann}\\ \\
Centre de Physique Th\'eorique CNRS\\
Case 907 Campus de Luminy\\
F-13288 Marseille Cedex 9\\
France\\ \\ \\
{\sc P. H. Damgaard$^a$, U. Magnea$^b$}\\ \\
$^a$The Niels Bohr Institute and $^b$NORDITA\\ 
Blegdamsvej 17\\ DK-2100 Copenhagen \O\\
Denmark\\ \\ \\
{\sc S. M. Nishigaki}\\ \\
Institute for Theoretical Physics\\
University of California\\
Santa Barbara, CA 93106-4030\\
USA}

\maketitle\vfill
\begin{abstract} We find the microscopic spectral densities and the
spectral correlators associated with multicritical behavior for both 
hermitian and complex matrix ensembles, and show their universality. 
We conjecture that microscopic spectral densities of Dirac operators 
in certain theories without spontaneous chiral symmetry breaking may 
belong to these new universality classes. 
\end{abstract}
\vfill
\begin{flushleft}
CPT-97/P.3526 \\
NBI-HE-97-45 \\
NORDITA-97/73\\
NSF-ITP-97-137\\
hep-th/9712006
\end{flushleft}
\thispagestyle{empty}
\newpage

\setcounter{page}{1}

\setcounter{equation}{0}
\section{Introduction}

\noi
It has been conjectured that the spectra of massless Dirac operators near
eigenvalues $\lambda = 0$ display universal features that can be
extracted from zero-dimensional theories of random matrices \cite{V,V1}.
The natural object to focus on is the so-called microscopic spectral
density
\beq
\rho_S(\lambda) ~\equiv~ \lim_{V\to\infty}\frac{1}{V}\rho\left(\frac{\lambda}
{V}\right) ~, \label{microdef}
\eeq
which displays the behavior of the spectral density near the origin,
as measured on the scale of the space-time volume $V$.

\noi
For QCD-like theories in $(2n+1)$-dimensions (gauge groups $SU(N_c)$
for $N_c \geq 3$, $N_f$ fermions in the fundamental representation
of the gauge group) the relevant matrix ensemble is that of
hermitian matrices, the unitary ensemble \cite{V1}. 
In even space-time dimensions, where chiral
symmetry can be defined, its possibly spontaneous breakdown
is known to be related to a non-vanishing spectral density at
the origin \cite{BC}: $\langle \bar{\psi}\psi\rangle \sim \rho(0)$.
In odd space-time dimensions the closest analogue to chiral symmetry is
parity. This raises the obvious question of whether flavor or flavor-parity
symmetries can be spontaneously broken in odd dimensions, similar
to the breaking of chiral flavor-symmetries in even dimensions.
While there is yet no direct evidence from, $e.g.$, Monte Carlo simulations
that this actually occurs
for QCD-like theories (see, however, ref. \cite{H}), 
one can entertain this idea, and look at some
of the consequences. It has been suggested that for an
even number of flavors, and three space-time dimensions, 
the pattern of symmetry breaking may be of
the flavor-breaking kind $U(N_f) \to U(N_f/2)\times U(N_f/2)$
\cite{P,V1}. 
An order parameter for such a transition is $|\Sigma| \equiv 
|\langle\bar{\psi}\psi\rangle|$,
in complete analogy with the breaking of chiral symmetries in even
dimensions. By the Banks-Casher relation \cite{BC}, this order
parameter is proportional to the spectral density at the origin,
$\rho(0)$. In such a scenario one can
evaluate the microscopic spectral density $\rho_S(\lambda)$ by
methods very analogous to the even-dimensional case \cite{V1}. 
The result has been found to be \cite{NS,V1}
\beqn 
\rho_S(\lambda) &=& \frac{1}{4}\pi^2\rho(0)^2 \lambda\left\{
J_{N_f+\frac{1}{2}}(\pi\rho(0)\lambda)^2 + J_{N_f-\frac{1}{2}}(\pi\rho(0)
\lambda)^2\right. \cr &&- \left.J_{N_f+\frac{1}{2}}(\pi\rho(0)\lambda)
J_{N_f-\frac{3}{2}}(\pi\rho(0)\lambda) - 
J_{N_f-\frac{1}{2}}(\pi\rho(0)\lambda)
J_{N_f+\frac{3}{2}}(\pi\rho(0)\lambda)\right\}\label{first}
\eeqn
for the Gaussian distribution. As was recently proven in ref. \cite{us}, 
this microscopic spectral density is highly
universal. It follows from hermitian random matrix models with arbitrary 
potentials $V(M)$ that support
a large-$N$ spectral density which is non-vanishing at the origin.
It was confirmed in \cite{V1} that this microscopic spectral density
is consistent with the
appropriate generalization of the Leutwyler-Smilga spectral sum rules 
\cite{LS} to this odd-dimensional situation.

\noi
The question of whether the symmetry breaking $U(N_f) \to U(N_f/2)\times
U(N_f/2)$ actually occurs in, say, (2+1)-dimensional QCD, is of course
a dynamical question that is not addressed in the above considerations.
To achieve such a spontaneous symmetry breaking, the eigenvalues of the 
Dirac operator must accumulate, as the volume $V$ is taken to infinity, 
sufficiently fast near $\lambda = 0$. A simple scaling argument shows
that the accumulation must be such that the average level spacing
$\Delta\lambda$ among the eigenvalues must be roughly constant near
$\lambda = 0$, and inversely proportional to the volume $V$. This is
precisely what motivates the introduction of the microscopic density
$\rho_S(\lambda)$ as in eq. (\ref{microdef}). An obvious question to
ask, then, is to what extent universal microscopic spectral
densities can be extracted if there is {\em no} spontaneous symmetry
breaking. In other words, suppose the dynamics of the massless theory 
is such that the 
eigenvalues of the Dirac operator do {\em not} accumulate sufficiently
fast for a condensate to form. Then the whole connection to the
Leutwyler-Smilga sum rules (and their odd-dimensional generalizations)
is lost, and the argument in favor of universal microscopic
spectral densities has to be reconsidered. This question is
particularly interesting in the case of (2+1)-dimensions, because there
the spontaneous breaking of flavor or flavor-parity symmetries is far from 
obvious. 

\noi
But also the even-dimensional case 
(corresponding to integration over complex matrices,
the so-called chiral unitary ensemble, \cite{TRM}
for QCD-like theories) merits closer attention in this
connection. For example, the universal results of refs.
\cite{V,us} are derived for an arbitrary number of flavors $N_f$.
The dependence on $N_f$ is simple, occurring only in the indices of the
relevant Bessel functions. This rather mild variation as a function
of $N_f$ is totally consistent with the fact that the only
essential flavor dependence enters in the precise chiral symmetry
breaking pattern $SU_L(N_f)\otimes SU_R(N_f) \to SU(N_f)$, and not
in detailed dynamical questions such as quark screening etc. Surely
the assumption about this chiral symmetry breaking pattern must break
down as the number of flavors is increased. For QCD an almost certain
upper limit is (for quarks in the fundamental representation) $N_f = 17$, 
for which the leading-order coefficient of the beta function has 
changed sign. The renormalized theory in this case is presumable ``trivial'',
and almost certainly not in a phase of chiral symmetry breaking.\footnote{
The number $N_f = 17$ at which chiral symmetry breaking no longer 
occurs is probably too conservative. Lattice gauge theory
simulations indicate that it may occur much before \cite{lattice}.}
However, although the renormalized coupling for infinite cut-off may
be vanishing, it is perhaps not {\em a priori} obvious that
the spectrum of the Dirac operator is 
identical to that of a totally free theory. So although the eigenvalues
may not accumulate fast enough near $\lambda = 0$ to produce a chiral
condensate, the accumulation rate may still be very different from that
of a theory with no bare interactions at all. This raises the question
whether also such spectral densities display universal features that are
computable from random matrix theory. As mentioned, it is even quite
likely that QCD possesses a large ``conformal window'' in a range of flavors,
where $N_f \leq 16$ but still greater than some critical value $N_f^*$. 
In such a
phase, which is not expected to support the spontaneous breaking of
chiral symmetry, the spectrum of the Dirac operator will definitely not
be that of a totally free theory either. Finally, although the scenario
is quite different in three dimensions (because the gauge theory is
superrenormalizable there, with a trivial beta function), the outcome
of adding more fermionic flavors to the theory is believed to be quite
similar. More flavors imply more screening of charges, and hence
reduced forces between the fermions. Eventually the forces should become
too small to cause chiral symmetry breaking. So even if flavor 
symmetries are spontaneously broken in three dimensions for a low number 
of fermionic flavors, one does not expect this to persist as $N_f$ 
is increased.

\noi
Yet another example may be general $SU(N_c)$ gauge theories coupled
to both fermions and Higgs fields. In the Higgs phases of such theories
we again do not expect chiral symmetry breaking. In the absence of
Yukawa couplings the Dirac operators of such theories are identical
to those without Higgs fields, and the arguments of ref. \cite{V}
would naturally lead one to associate analogous matrix models {\em
with vanishing spectral density at the origin} to such phases. 
In fact, if one mentally integrates out the Higgs fields,
the resulting theories are formally of the same class as the theories
without Higgs fields. Their only difference is that they belong to
different phases. A totally different class of gauge theories in
four dimensions which do not cause spontaneous chiral symmetry breaking
are those of Abelian gauge groups, such as QED. Also for these much
simpler theories one can ask whether their Dirac operator spectrum
near the origin contain universal functions.

\noi
All of these physical considerations have led us to analyze the fate of the
universal eigenvalue correlators (and in particular their microscopic
spectral densities) when the the macroscopic spectral densities at the
origin reach zero, {\em i.e.} when $\rho(0) = 0$ but without a finite 
gap in eigenvalues around $\lambda = 0$. {}From a pure matrix
model point of view, these questions are clearly posed, and may be considered
in their own right. For example the proof for the  macroscopic universality
of density-correlators with such a multi-band support for the eigenvalues
has been recently given for the hermitian and complex matrix model
\cite{AA}.
Their possible physical relevance in
terms of applications to the Dirac operator spectrum near the origin
can necessarily only be phrased in terms of conjectures. They are,
however, highly non-trivial statements, and they are directly testable
in, for example, lattice gauge theory simulations.\footnote{For a 
beautiful comparison between recent lattice gauge theory data and 
the universal predictions from matrix models in the usual case of 
chiral symmetry breaking, see ref. \cite{BBMSVW}.}

\noi
To require that the (macroscopic) spectral density precisely just vanishes 
at the origin puts a constraint on the matrix model in question. This
implies the fine tuning of at least one parameter (coupling) in the
matrix model potential $V(M)$. {}From this point of view it is not
surprising that if we associate the usual scaling (with no such constraint) 
with normal critical behavior, the just-vanishing of the spectral
density at the origin corresponds to, in general, multicritical behavior.
There are, for similar reasons, tight constraints on the precise manner
in which the macroscopic spectral density can approach zero at the origin,
in general labeled by an integer $m$, which counts the order of 
multicriticality. To achieve such multicriticality of $m$th order, 
$m$ couplings must be tuned.
This integer $m$ in turn describes the rate at which
eigenvalues accumulate near the origin, a rate which hence cannot be totally
arbitrary. 

\noi
With these preliminary remarks to serve as motivation, we now turn to
the details of the actual matrix model calculations. Our analysis will
be based on the orthogonal polynomial technique, which is known to
greatly simplify many calculations in the unitary and chiral unitary 
matrix ensembles. As in ref. \cite{us}, we shall turn the three-step
recursion relations for these orthogonal polynomials into a differential
equation in the large-$N$ limit. {}From this we can derive the limiting
forms of the orthogonal polynomials in the appropriate large-$N$ scaling
regions. A recent paper by Kanzieper and Freilikher \cite{KF} shows 
very elegantly how the derivation of this differential equation can
be simplified. Their approach is in fact more powerful, and leads more
easily to generalizations. We shall essentially follow their approach here.
As derived in ref. \cite{KF}, the crucial
property of the differential equation is that it does not explicitly depend 
on the chosen matrix model potential $V(M)$ once the large-$N$ limit is 
taken, but only on the associated {\em macroscopic} spectral density 
$\rho(\lambda)$. As we shall see in section 2, this simplifying
feature unfortunately does not survive in the particular microscopic
large-$N$ limit that is required to probe the multicritical domain
near $\lambda \!=\! 0$. The resulting differential equation will therefore
not manifestly depend only on the macroscopic spectral density 
$\rho(\lambda)$, but will also contain contributions from what are naively 
subdominant terms in the $1/N$ expansion. Universality of the orthogonal 
polynomials, and hence universality of the microscopic spectral density and 
the spectral correlators, is hence not as obvious as in the usual case.
Finding those naively subleading corrections that turn out to contribute
in the microscopic multicritical limit entails solving an auxiliary
differential equation, known in a different context as the string 
equation.\footnote{Only after properly solving this 
additional differential equation, and
inserting its solution, does the differential equation for the orthogonal
polynomials become manifestly universal.}
For the
case at hand it turns out to be a generalized Painlev\'{e} II equation, 
which we show how to solve accurately by numerical methods. Both because
the solution to the generalized Painlev\'{e} II equation cannot be found
in closed analytical form, and because the underlying differential
equation for the orthogonal polynomials themselves cannot be recast
in a particularly simple form, it is not possible to give a closed analytical
expression for the microscopic spectral correlators in this 
multicritical case. The solution can be found to any required
accuracy by numerical means. We do however identify an intermediate
``mesoscopic'' range at which the spectral correlators can be
computed analytically, expressed by some combination of Bessel functions.
In section 3 we derive, for completeness, the analogous results for
the chiral-unitary matrix ensembles. They may be relevant for Dirac
operators of fermions in the fundamental representations of gauge groups
$SU(N_c\!\geq\!3)$ in even space-time dimensions. These results follow
as simple corollaries from the unitary ensemble. Section 4 contains our
conclusions.

\setcounter{equation}{0}
\section{Multicritical Unitary Ensembles}

We begin with the unitary matrix ensemble described by the partition
function
\beqn
{\cal Z} &=& \int\!dM {\det}^{2\alpha}M ~e^{-N Tr V(M)} \nonumber\\
       &\sim& \int_{-\infty}^{\infty}\! \prod_{i=1}^N \left(d\lambda_i
|\lambda_i|^{2\alpha}e^{-NV(\lambda_i)}\right)\left|{\det}_{ij}
\lambda_j^{i-1}\right|^2 ~,
\label{Z}
\eeqn
with some even potential
\beq
V(M) ~=~ \sum \frac{g_{2k}}{2k} M^{2k} ~.
\eeq
The matrices $M$ are $N\times N$ and hermitian, with real eigenvalues $\la_i$. 
The integration is over
their associated Haar measure $dM$. This matrix model is, for generic
potentials $V(M)$ giving rise to a single cut containing the origin
of eigenvalues $\lambda\! =\! 0$, conjectured to be related to the
Dirac operator spectrum of $SU(N_c\!\geq\! 3)$ gauge theories coupled
to an even number of flavors $N_f\! =\! 2\alpha$. For that application
$\alpha$ is thus restricted to be an integer. Using the formalism
of Kanzieper and Freilikher \cite{KF1} (see also refs. \cite{SBC,TW}) 
this restriction can be lifted without difficulty,
and we shall therefore in greater generality consider any real 
$\alpha > -1/2$. We shall largely employ the notation of ref. \cite{KF1}
in what follows below. 

\noi
Using the method of orthogonal polynomials to analyze the above partition
function, one has to find polynomials $P_n(\lambda)$
orthogonal with respect to the weight
function $w(\lambda) \equiv |\lambda|^{2\alpha}\exp[-NV(\lambda)]$. We 
choose them orthonormalized:
\beq
\int_{-\infty}^{\infty}\! d\lambda~ w(\lambda) P_m(\lambda)P_n(\lambda)
~=~ \delta_{mn} ~. \label{ONP}
\eeq
It is actually more convenient to work directly with the
``wave functions''
\beq
\psi_m(\lambda) ~\equiv~ |\lambda|^{\alpha}
e^{-\left(\frac{N}{2}\right)V(\lambda)}P_m(\lambda) 
~,
\l{defpsi}
\eeq
because these are the objects that enter directly into the determination
of the spectral kernel (two-point function):
\begin{eqnarray}
K_N(\lambda,\mu) &~=~& \sum_{n=0}^{N-1}\psi_n(\lambda)\psi_n(\mu) \cr
&~=~& c_N \frac{\psi_N(\mu)\psi_{N-1}(\lambda)
- \psi_N(\lambda)\psi_{N-1}(\mu)}{\mu-\lambda} ~.
\end{eqnarray}
Here the coefficients $c_N$ are those of the three-step recursion
relation for the orthogonal polynomials:
\beq
\lambda P_{n-1}(\lambda) ~=~ c_nP_n(\lambda) + c_{n-1}P_{n-2}(\lambda) ~.
\label{rec}
\eeq
They can be determined from the following identity\footnote{By switching 
to the monic normalization $\tilde{P}_n(\la)=\la^n +\ldots$ one can easily 
derive that $P_n^{~\prime}(\lambda)=\frac{n}{c_n}P_{n-1}(\lambda)
+{\cal O}(\la^{n-2})$.} 
(which in the context of 2-d quantum gravity is called the string equation) 
by partial integration, using eq. (\ref{rec}),
\beq
n ~=~ c_n 
\int_{-\infty}^{\infty}\! d\lambda~ w(\lambda) P_n^{~\prime}(\lambda)
      P_{n-1}(\lambda) ~.
\l{string}
\eeq
The wave functions $\psi_m(\lambda)$ are clearly orthonormalized on the
usual (flat) Lebesgue measure.

\noi
In order to derive a differential equation for the wave functions it is 
convenient to express the differential operator acting on the orthogonal
polynomials in the following way \cite{KF1}:
\beq
P_n^{~\prime}(\lambda) ~\equiv~ A_n(\lambda)P_{n-1}(\lambda) ~-~
                       B_n(\lambda)P_{n}(\lambda) ~,
\label{Pprime}
\eeq
where the functions $A_n$ and $B_n$ have to be determined.
Following Kanzieper and Freilikher \cite{KF1}, they are given by
\begin{eqnarray}
A_n(\lambda)&=&N~ c_n\int_{-\infty}^{\infty}\! d\mu~ \frac{w(\mu)}{\mu-\lambda}
     [V_{(\alpha)}'(\mu) - V_{(\alpha)}'(\lambda)]P_n(\mu)^2 ~,\label{An}\\
B_n(\lambda)&=&N~ c_n\int_{-\infty}^{\infty}\! d\mu~ \frac{w(\mu)}{\mu-\lambda}
  [V_{(\alpha)}'(\mu) - V_{(\alpha)}'(\lambda)]P_n(\mu)P_{n-1}(\mu) \nonumber\\
         &\equiv& B_n^{\mbox{\scriptsize reg}}(\lambda)
            ~+~ (1-(-1)^n)\frac{\alpha}{\lambda} ~, \label{Bn}
\end{eqnarray} 
where the $\alpha$-dependent part of the measure has been absorbed into
the potential\footnote{In $A_n(\la)$ the term proportional to $\alpha$
vanishes due to the potential $V_{(\alpha)}(\lambda)$ being even, whereas for 
$B_n(\la)$ it is given explicitly, following again \cite{KF1}.}: 
\beq
V_{(\alpha)}(\lambda) ~\equiv~ V(\lambda) - \frac{2\alpha}{N}\ln|\lambda| ~. 
\l{Valpha}
\eeq
In contrast to reference \cite{KF1} we have chosen the usual matrix model
convention eq. (\ref{Z}) which makes the dependence on $N$ explicit. 
The evaluation of different powers in $N$ will become very important in the 
multicritical 
case. Using the results eqs. (\ref{An}) and (\ref{Bn}) as well as the 
recursion relation eq. (\ref{rec}) the following identity is exactly satisfied 
\cite{KF1}:
\beq
B_n(\lambda) + B_{n-1}(\lambda) ~+~  N~ V_{(\alpha)}'(\lambda) 
~=~ \frac{\lambda}{c_{n-1}}A_{n-1}(\lambda)~.
\label{ABV}
\eeq

\noi
Differentiating eq. (\ref{Pprime}) for $\psi_n(\lambda)$ and using relations
(\ref{rec}) and (\ref{ABV}) 
finally leads to the following exact differential equation for
the wave functions $\psi_n(\lambda)$, valid for any $n$ \cite{KF1}: 
\beq
\psi_n''(\lambda) - F_n(\lambda)\psi_n'(\lambda) + G_n(\lambda)
\psi_n(\lambda) ~=~ 0 ~,
\label{diff}
\eeq
where
\begin{eqnarray}
F_n(\lambda) &\equiv& \frac{A_n'(\lambda)}{A_n(\lambda)} \l{KFF}\\
G_n(\lambda) &\equiv& \frac{c_n}{c_{n-1}}A_n(\lambda)A_{n-1}(\lambda) 
                -\left(B_n(\lambda)+\frac{N}{2}V_{(\alpha)}'(\lambda)\right)^2
               \nonumber\\
                &&+\left(B_n(\lambda)+\frac{N}{2}V_{(\alpha)}'(\lambda)\right)'
                -\frac{A_n'(\lambda)}{A_n(\lambda)}
                \left(B_n(\lambda)+\frac{N}{2}V_{(\alpha)}'(\lambda)\right) ~.
\l{KFsGF}
\end{eqnarray}

\noi 
The rest of our paper will consist in analyzing the way in which
suitable microscopic large-$N$ limits affect the above differential
equation. Performing the usual large-$N$ limit the recursion coefficients
$c_n$ and the functions $A_n(\lambda)$ and 
$B_n^{\mbox{\scriptsize reg}}(\lambda)$ are 
assumed to 
approach smooth functions. Under these conditions,
with a generic potential $V(\lambda)$, the function $A_n(\lambda)$
is directly related to the 
macroscopic spectral density $\rho(\lambda)$ \cite{KF,SNtalk}:
\beq
\rho(\lambda) ~=~ \lim_{N\to\infty}\frac{1}{N\pi}
    A_N(\lambda)\sqrt{1 - (\lambda/a)^2} ~,
\label{Arho}
\eeq
where $a\!=\! \lim_{N\to\infty}c_N$ is the endpoint of the spectrum. 
Consequently, because of eq. (\ref{ABV}), the asymptotic differential equation
(\ref{diff}) for $\psi_N(\lambda)$ should only depend on the macroscopic
spectral density. But precisely at multicriticality,
and in the appropriate microscopical scaling limit, the simple relationship
(\ref{Arho}) unfortunately breaks down. This is due to the contribution of 
terms that are naively subleading in $1/N$. Because of this, the appropriate 
differential equation for the wave functions $\psi_N(\lambda)$ in the 
multicritical microscopic limit will {\em not} just depend on the spectral 
density $\rho(\lambda)$. To see this, we first make the definition of
multicriticality precise.

\vspace{0.2cm}
\subsection{\sc Multicritical potentials}\l{multic}

Multicriticality can be reached at both the endpoint of the spectrum,
and at the origin. Because of the physical motivations mentioned in
the introduction, we here focus exclusively on multicriticality at the
origin of the spectrum. The case of $\rho(0) \neq 0$ will thus play,
in this context, the r\^{o}le of being uncritical ($m\!=\!0$). 
As $\rho(0) \to 0$, the one-cut solution around the origin
will turn into a two-cut solution, thereby defining the
first criticality ($m\!=\!1$). As is well known, this happens
in the simplest case \cite{S,CMM,DSS} when the potential $V(\lambda)$ has
a double-well shape, as shown in fig.~1.\\

\noindent
\begin{minipage}{\textwidth}
\begin{center}
 \leavevmode
 \figinclude{237}{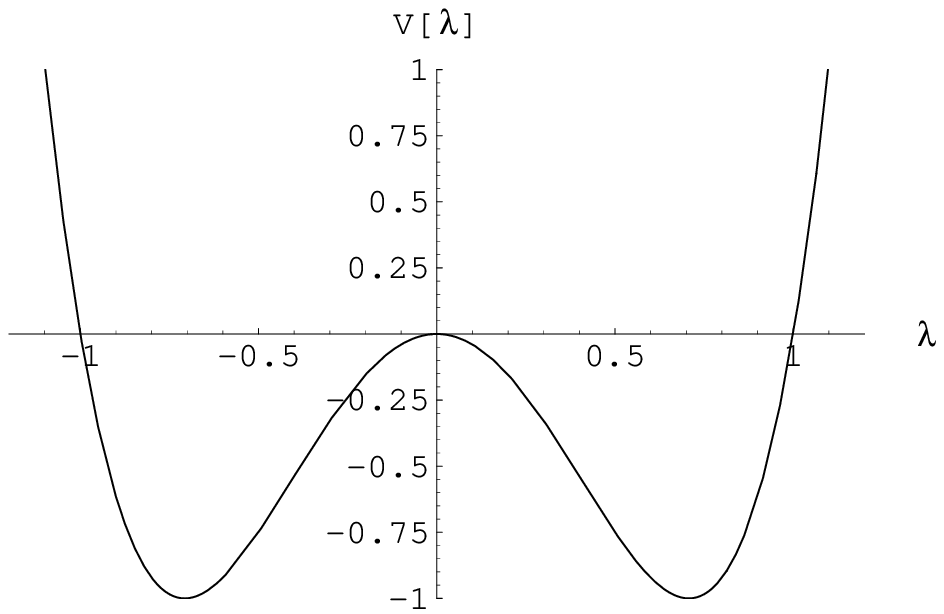}
 \figinclude{237}{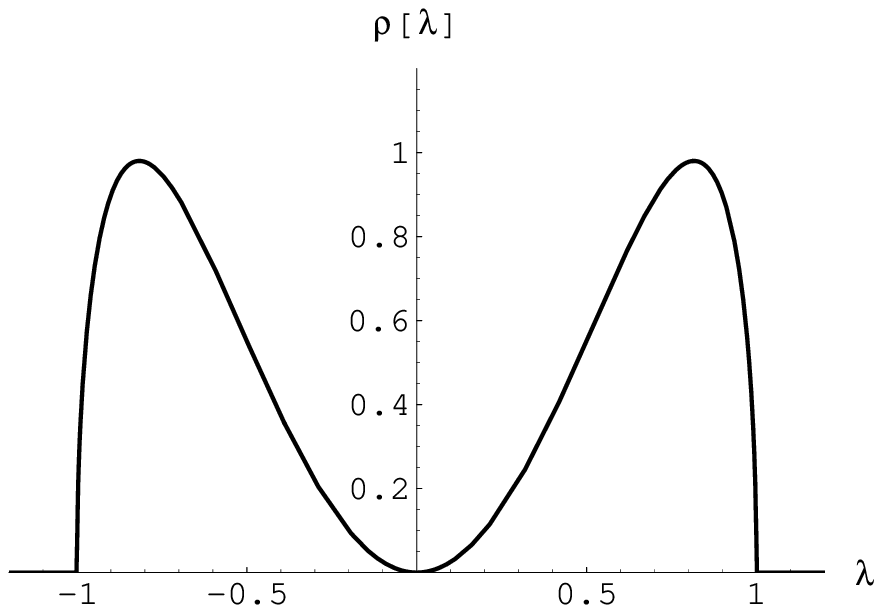}\\
\end{center}
\centerline{{\small Figure 1: The critical $m\!=\! 1$ potential (left) and
the associated spectral density (right).}}
\end{minipage}

\noi
A characterization of potentials leading to multicriticality at 
the origin has been given by Crnkovi\'c and Moore \cite{CM}. In general, for
polynomial potentials of degree $2k$, 
the macroscopic spectral density is, for a symmetric one-cut solution, 
of the form
\beq
\rho(\lambda) ~=~ Q_{2k-2}(\lambda)\sqrt{1 - (\lambda/a)^2} ~,
\eeq
where $Q_k(\lambda)$ is a polynomial of degree $k$. Generically, $Q_k(0)
\neq 0$, in which case $\rho(\lambda) \sim$ const. near $\lambda\!=\!0$. 
To force $\rho(0) \to 0$ thus entails tuning the potential $V(\lambda)$
so that the polynomial $Q(\lambda)$ vanishes at the origin.
Because $Q(\lambda)$ is a polynomial of even degree, this can be achieved
by powers of an even integer, $2m$, where $m$ labels the order of
multicriticality:
\beq
\rho(\lambda) ~=~ \mbox{\rm const.}~ \lambda^{2m} + ~\ldots ~, ~~~~~~~~~~~
{\mbox{\rm as}}~~ \lambda \to 0 ~.\label{rhomc}
\eeq
Polynomial potentials $V_m(\lambda)$ that lead to such behavior are
thus at least of degree $2m\!+\!2$.
A particular class of such potentials for which the
behavior of the macroscopic spectral density is as above can be
compactly described by \cite{CM}\footnote{Crnkovi\'c and Moore have considered
a spectral density with two symmetric cuts merging. This leads to the the
same multicritical potentials $V_m(\la)$ as a symmetric one-cut density
developing zeros at the origin.}
\beq
V_m'(\lambda) ~=~ k(m)\lambda^{2m+1}
\left.\left(1 - \frac{1}{\lambda^2}\right)^{1/2}\right|_+ ~,\label{vprime}
\eeq
where endpoint of the support $a$ has been set to unity and
the ``+'' subscript indicates that only the polynomial part in
an expansion around $\lambda\!=\!\infty$ is kept. The constant in front,
chosen so as to normalize the macroscopic spectral density by
\beq
\int_{-1}^{1}\! d\lambda \rho(\lambda) ~=~ 1
\eeq
is given by 
\beq
k(m) ~=~ 2^{2m+1}\frac{(m+1)!(m-1)!}{(2m-1)!} ~.
\eeq
The spectral density at the multicritical point is then given by
\beq
\rho_m(\la) ~=~ \frac{1}{2\pi} k(m) \la^{2m} \sqrt{1-\la^2} ~.
\label{rhocrit}
\eeq
Conforming with common usage, we shall denote such multicritical potentials
as ``minimal''. An infinity of other potentials can be constructed
by adding higher powers of $\la$ the polynomial $Q_k(\la)$, 
which lead to precisely the same multicritical behavior (\ref{rhomc}). For
this reason one can speak about an $m$th multicritical universality class. 
One sees immediately that the $\alpha$-dependent term in our integration
measure has no influence on the classification of these multicritical
potentials, since this term is suppressed by one power of $1/N$ in the
planar limit (see eq. (\r{Valpha})).

\vspace{0.2cm}
\subsection{\sc The breakdown of the usual planar limit}

As it has been mentioned earlier in this section an asymptotic differential 
equation for $\psi_N(\la)$ at multicriticality is no longer easy to obtain
due to a breakdown of the relation (\r{Arho}). In order to show this let us 
briefly review how the result is obtained in the noncritical case, where the
usual planar limit produces all relevant terms, and point out what breaks down
at multicriticality. 

\noi
In the usual planar limit the following behavior is assumed for the recursion
coefficients:
\beq
c_{n\pm 1} ~\rar~ c_n + {\cal O}(1/N) ~,
\l{clim}
\eeq
which is directly related to the fact that we are here considering a 
one-cut phase. Moreover, an implicit assumption in ref. \cite{KF1} is that
the following smooth limits are satisfied:
\beqn
A_{n\pm 1}(\la) &\rar& A_n(\la) + {\cal O}(1/N) ~, \l{Alim} \\
B_{n\pm 1}^{\mbox{\scriptsize reg}}(\la)  
    &\rar& B_n^{\mbox{\scriptsize reg}}(\la) + {\cal O}(1/N) ~. \l{Blim} 
\eeqn
The requirement eq. (\r{clim}) together with the leading term of the
relation
\beq
2c_N ~=~ a + \ldots
\eeq
between the coefficient $c_N$ and 
the endpoint of the support leads to the asymptotic relation (\r{Arho})
between $A_N(\la)$ and the macroscopic spectral density $\rho(\la)$.
Inserting the limit (\r{Blim}) into eq. (\r{ABV}) leads asymptotically to
\beq
B_N(\la) + \frac{N}{2}V_{(\alpha)}'(\la) ~\rar~ \frac{\la}{2c_N}A_N(\la)-(-1)^N
            \frac{\alpha}{\la} ~.
\eeq
Consequently $G_N(\la)$ can be expressed asymptotically by $A_N(\la)$
(and via eq. (\r{Arho}), if true, by $\rho(\la)$) 
only\footnote{The last term has been omitted in eq. (34) of ref. \cite{KF1} 
because the authors did not consider the microscopic limit at the origin.}.
\beq
G_N(\la) ~\rar~ A_N^2(1-\frac{\la^2}{a^2})+\frac{(-1)^N\alpha-\alpha^2}{\la^2}
                +(-1)^N\frac{\alpha}{\la}\frac{A_N'(\la)}{A_N(\la)} ~.
\l{Gnaiv}
\eeq
Looking at the noncritical case at the origin the terms proportional to 
$\frac{A_N'(\la)}{A_N(\la)}$ are subleading in the microscopic limit at 
the origin so the last term in $G_N(\la)$
and the $F_N(\la)$-term can be disregarded. This leads to the asymptotic
differential equation
\beq
\psi_N''(x) + \left( \frac{A_N(0)^2}{N^2} 
        +\frac{(-1)^N\alpha-\alpha^2}{x^2}\right) \psi_N(x) ~=~ 0 ~,
\l{diffold}
\eeq
where the appropriate scaling $x\! =\! N\la$ has been used
and $\psi_N(\la=\frac{x}{N})$ is regarded a function of $x$.
It has been derived and solved for nonnegative integers in \cite{us},
and generalized to real $\alpha>-1/2$ in \cite{KF1}.
Apart from the independence of the two derivations, we have also checked
the resulting solutions (certain Bessel functions, see ref. \cite{us}) 
by numerically determining $\psi_N(\la)$ directly from the recursion
relations themselves, for very large values of $N$.

\noi
Moreover there is an analytic consistency check on the differential equation
(\r{diffold}). It is straightforward to verify (see Appendix A) that
the following exact symmetry holds for the wave functions:
\beq
\psi_{2n}^{(\alpha+1)}(\la) ~=~\mbox{sign}(\la)\psi_{2n+1}^{(\alpha)}(\la) ~.
\l{asym}
\eeq
The superscript indicates the corresponding value of $\alpha$ in the measure.
We will refer to this symmetry as $\alpha$-symmetry. Note that the
macroscopic spectral density is independent of $\alpha$, since the
$\alpha$-dependence can be considered as a subleading term in the
potential. Due to relation (\r{Arho}) and the limit (\r{Alim}) the 
differential equation (\r{diffold}) is therefore manifestly invariant under 
the $\alpha$-symmetry. Indeed, the solutions of ref. \cite{us} trivially
satisfy the relation (\r{asym}).

\noi
Let us now turn to the multicritical case. The appropriate rescaling of
eigenvalues that is needed to obtain a non-trivial microscopic limit
will now be of the form
\beq
x ~=~ N^{\frac{1}{2m+1}}\la \l{mresc}
\eeq 
at the $m$th multicritical point ($m \in$ {\bf N}). 
We first suppose that the usual
planar limit eqs. (\r{clim}) - (\r{Blim}) still holds. One might object that 
the recursion coefficients now obey a period-two ansatz \cite{CMM} instead of
eq. (\r{clim}). But in the planar limit, and just at the critical point $n=N$,
the two branches merge to the one-cut value $c_N=a/2$, as we are right on top 
of the one-cut--two-cut transition. Therefore the argument in \cite{KF1} 
relating $A_N(\la)$ and $\rho(\la)$ (eq. (\r{Arho})) should still 
be expected to hold. Next, we have seen
from the previous subsection that at the $m$th multicritical point $\rho(\la)$
behaves like $\la^{2m}$ at $\la=0$. Consequently the last term in eq. 
(\r{Gnaiv}) for $G_N(\la)$ becomes $(-)^N2m\alpha/\la^2$ which is of the same 
(leading) order as the second term. 

\noi
So from the usual planar limit we are left with the full $G_N(\la)$ eq. 
(\r{Gnaiv}) in the asymptotic differential equation for $\psi_N(\la)$.
However, this differential equation is no longer invariant under the 
$\alpha$-symmetry. It therefore cannot be the correct equation. 
In order to confirm this, we have compared the 
prediction from this naive differential equation to the polynomials 
constructed by Gramm-Schmidt orthonormalization.
The mismatch is shown in fig.~2.\\

\noindent
\begin{minipage}{\textwidth}
\begin{center}
 \leavevmode
 \figinclude{237}{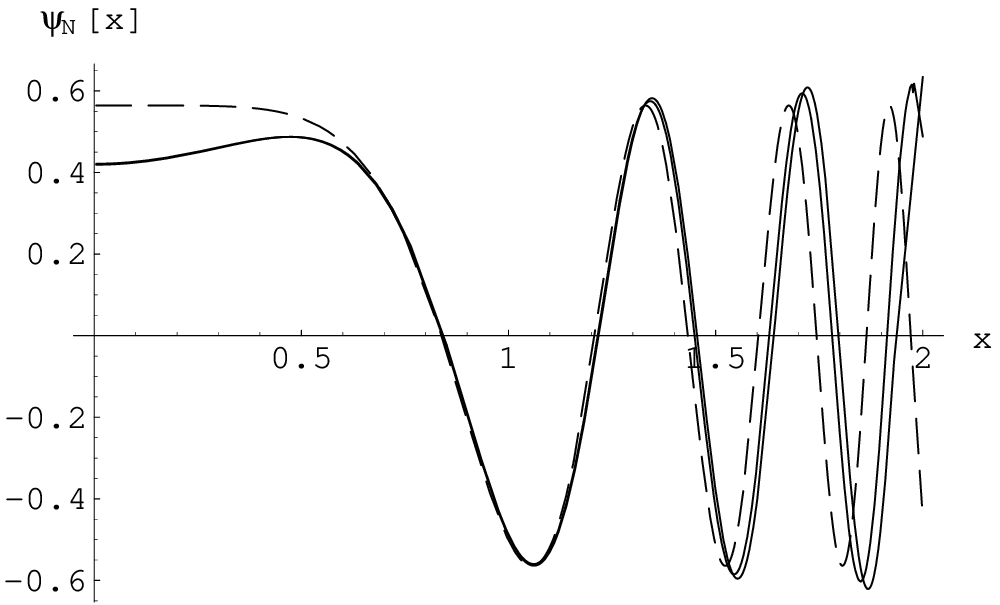}
 \figinclude{237}{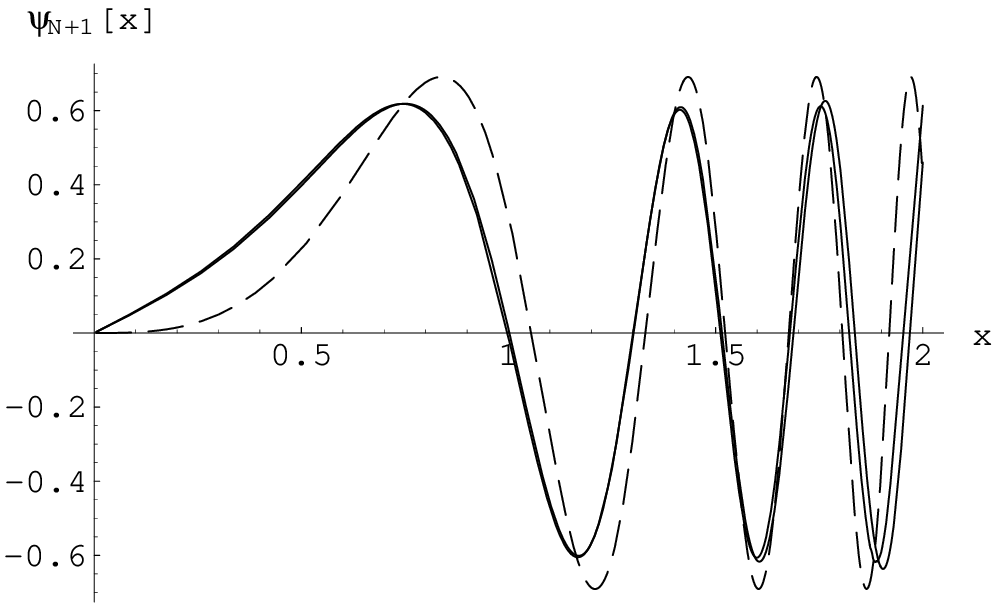}\\

\end{center}
\noindent 
{\small 
Figure 2:
The microscopic wave functions for the $m=1$ critical potential
(\r{vprime}) are plotted for $N=32$ and $48$ ($\alpha=0$) 
(real lines).
They converge to limiting functions which differ
from the solution of the naive differential equation 
(\r{Gnaiv}) (dotted lines) for small $x$.}
\end{minipage}

\noi
The only explanation for this failure is that the simple relationship
(\r{Arho}) breaks down or, in other words, that the $\alpha$-dependent
corrections to the recursion coefficients that were subleading before have 
to be taken into account. Remarkably, this can also be established directly
by analyzing the ``string equation'' for the coefficients $c_n$, as we will
discuss in section 2.4. One finds that corrections to the $c_n$'s 
are precisely of the order $N^{-\nu}$ for $\nu\! =\! 1/(2m+1)$ at 
$m$th multicriticality,
instead of $N^{-1}$ (as for the case $m=0$). It is due to this rather devious
compensation that the naive differential equation for the wave functions
is incorrect.

\noi
Despite many efforts we have not been able to derive a general expression
for the asymptotic differential equation containing all relevant contributions
from the $c_n$'s. The analytical formulas increase dramatically in
complexity with increasing $m$. We will therefore restrict ourselves to a 
detailed analytical treatment of the $m\!=\! 1$ and $m\!=\! 2$ 
multicriticalities. Many aspects of the formalism do however readily
generalize to arbitrary $m$, and we shall keep the discussion in as general
form as possible. In Appendix B we write down a detailed conjecture for the
differential equation for the wave functions at any multicriticality.

\vspace{0.2cm}
\subsection{\sc An example: the case $m=1$}\l{m=1expl}

In order to treat all contributions to the differential equation at
multicriticality we will restrict ourselves first to the simplest example,
the case $m\!=\!1$. This corresponds to a
minimal potential of the simple Mexican-hat form
\beq
V_1(\lambda) ~=~ \frac{1}{2}g_2\la^2 +\frac{1}{4}g_4\la^4 ~~~~\mbox{with}~~~~
g_2=-8 ~,~~g_4=16 ~.
\l{V1}
\eeq
The corresponding eigenvalue density reads\footnote{Here $a=1$ due to 
our conventions from subsect. \r{multic}.} 
\beq
\rho_1(\la) ~=~ \frac{1}{2\pi}g_4\la^2\sqrt{1-\la^2} ~.
\eeq
We are now able to give exact expressions for the functions $A_n(\la)$ and
$B_n(\la)$ only in terms of the recursion coefficients $c_n$. This will allow
us to keep track where the correction to the $c_n$'s will enter. We will then 
be able to derive an asymptotic differential equation at $m=1$ 
multicriticality. 

\noi
Making use of the completeness of the orthogonal polynomials we can explicitly
express $P_n^{~\prime}(\lambda)$ by lower order polynomials, as we have chosen
a particular potential:
\beqn
P_n^{~\prime}(\lambda) &=& \sum_{k=0}^{n-1}P_k(\lambda)
\int_{-\infty}^{\infty}\! d\mu~ w(\mu) N V_{(\alpha)}'(\mu) ~P_n(\mu)P_k(\mu)
\nonumber\\
&=& N c_n \left( g_2 + g_4(\cnp +\cn +\cnm )\right) P_{n-1}(\la) +
Nc_n c_{n-1} c_{n-2} g_4 P_{n-3}(\la) - (1-(-1)^n)\frac{\alpha}{\la}P_n(\la) 
\nonumber\\
\l{Ppr}
\eeqn
where the recursion relation (\r{rec}) has been used for the potential eq.
(\r{V1}). Looking back at the definition (\r{Ppr}) $A_n(\la)$ must be an even
function whereas $B_n^{\mbox{\scriptsize reg}}(\la)$ 
is odd. A simple way of finding them is to assume that they are 
polynomial\footnote{Remember that in the usual planar limit $A_N(\la)$ becomes
proportional to the polynomial $Q_k(\la)$ in front of the square root of 
$\rho(\la)$.}, use the recursion relation and then comparing coefficients. 
In this way one finds
\beqn
A_n(\la) &=& Nc_n \left( g_2 + g_4(\cnp +\cn) ~+~ g_4\la^2\right) 
\l{Acrit} \\
B_n(\la) &=& N\cn g_4 \la ~+~ (1-(-1)^n)\frac{\alpha}{\la} ~.
\l{Bcrit}
\eeqn
One can also derive these expressions directly, using the the definitions
(\r{An}) and (\r{Bn}). 
If we would now set $\cnp\! =\!\cn\! =\! 1/4$ and use the critical couplings
eq. (\r{V1}) the combination $g_2+g_4(\cnp +\cn)$ would vanish and we would
be back at the relation (\r{Arho}), which no longer holds at multicriticality.
{}From our example we are now in a position to determine the order 
and the explicit value of the corrections. Inserting the appropriate
microscopic scaling limit at multicriticality, eq. (\r{mresc}),
the whole differential equation (\r{diff}) has to be rescaled by
$N^{-\frac{2}{2m+1}}$, which is $N^{-2/3}$ in our case. In order to get
contributing corrections in the differential equation, the following terms
have to be of order one:
\beqn
u_{\pm} &\equiv& N^{\frac{2}{3}}\left( g_2 + g_4(c_{N\pm 1}^2+c_N^2)\right)
           \nonumber\\
v~ &\equiv& N^{\frac{1}{3}} \left( g_2 + 2g_4c_N^2\right) ~.
\l{uv}
\eeqn
In other words, for $m\!=\! 1$ the terms in the parenthesis must vanish 
like $N^{-2/3}$ and $N^{-1/3}$, respectively. This is precisely what we get 
from eqs. (\r{moorexp}) and (\r{sumcs}) below. 
Expressed in terms of the above quantities, we get the following
exact differential equation for the potential eq. (\r{V1}) 
(from eqs. (\r{Acrit}), (\r{Bcrit}) and (\r{KFsGF})):
\beq
\psi_N''(x) - N^{-1/3}F_N(x)\psi_N'(x) + N^{-2/3}G_N(x) \psi_N(x) ~=~ 0 ~.
\l{diffx}
\eeq
Here
\beqn
N^{-1/3}F_N(x) &=& \frac{2g_4x}{u_+ +g_4x^2} \nonumber\\
N^{-2/3}G_N(x) &=& c_N^2u_+u_- +\Big((-1)^N\alpha +\frac{1}{2}\Big) v 
 + \left( c_N^2(u_+ + u_-)g_4 -\frac{1}{4}v^2 +  N^{-1/3}\Big((-1)^N\alpha
  + \frac{3}{2}\Big) g_4\right)x^2 \nonumber\\
 &&+ \left( c_N^2g_4^2 -\frac{1}{2}N^{-1/3}vg_4\right)x^4 
    -\frac{1}{4}N^{-2/3}g_4^2x^6
+\frac{(-1)^N2g_4\alpha -vg_4x^2 - N^{-1/3}g_4^2x^4}{u_+ +g_4 x^2}\nonumber\\ 
 && +~\frac{(-1)^N\alpha-\alpha^2}{x^2} ~.
\l{FG}
\eeqn
Due to eq. (\r{uv}) it is now clear that the terms being explicitly 
proportional to negative powers of $N$ will vanish in the microscopic large-$N$
limit. Moreover, the dominant term $\sim x^2$ in $G_N(x)$ vanishes up to
${\cal O}(N^{-1/3})$:
\beq
v^2 ~=~ 4c_N^2g_4 (u_+ +u_-) ~.\l{uvid}
\eeq
This follows directly from the string equation (\r{string}) at $n=N$, 
which in our example reads
\beq
\frac{n}{N} +(1-(-1)^n)\frac{\alpha}{N}~=~ 
c_n^2 \left( g_2 + g_4(\cnp +\cn +\cnm)\right) ~,
\l{stringexpl}
\eeq
and from inserting the critical values for the coupling constants.
Finally the correct asymptotic differential equation for $m=1$ criticality
is eq. (\r{diffx}) with 
\beq
N^{-2/3}G_N(x) ~=~ c_N^2u_+u_- +\left((-1)^N\alpha -\frac{1}{2} \right)v ~+~ 
\frac{u_+v+(-1)^N2\alpha g_4}{u_+ + g_4x^2} ~+~ c_N^2g_4^2x^4 ~+~
\frac{(-1)^N\alpha -\alpha^2}{x^2}~,
\l{Gasympt}
\eeq
where one of the quantities $u_+,u_-$ or $v$ can still be eliminated by
means of eq. (\r{uvid}). After having determined the functional form of the
differential equation, which is now no longer a Bessel equation,
the only point left to be done is the analytical calculation
of the corrections to, say, $\cnp$ and $\cn$. This will be the topic of 
the next subsection.

\noi
Let us add a final remark on the $\alpha$-symmetry of this new asymptotic
differential equation. The function $F_n(\la)$ can be seen to be invariant
for finite $n$ from the definition eq. (\r{KFF}) since one has
\beq
\frac{A_{2n}^{(\alpha+1)}(\la) }{c_{2n}^{(\alpha +1)}}~=~
\frac{A_{2n+1}^{(\alpha)}(\la) }{c_{2n+1}^{(\alpha )}} ~.
\eeq
This follows directly from eq. (\r{asym}). An immediate consequence 
of this identity is the invariance of $u_+$ under the $\alpha$-symmetry.
Still, apart from the last two terms, $G_N(x)$ is not
obviously invariant. In order to prove its invariance under the
$\alpha$-symmetry 
one would have to know precisely the behavior of $u_-$ and $v$ as functions
of $\alpha$. As we shall see below, this is a highly non-trivial problem,
related to the solution, for arbitrary $\alpha$, of a certain non-linear
differential equation.

\vspace{0.2cm}
\subsection{\sc Painlev\'{e} II}

The apparently subdominant terms that enter the differential 
equation for the wave functions $\psi(\lambda)$ in the microscopic limit near
multicriticality arise from corrections to the coefficients $c_n$ 
as $N\!\to\!\infty$. It is well-known \cite{CMM,DSS} that in the case 
at hand the
coefficients $c_n$ do not approach one single smooth function in the
large-$N$ limit, but rather split into two functions depending on whether
the index is odd or even. In all generality \cite{CM}\footnote{Here we have
used the most general ansatz for the coefficients $c_n$.}
\begin{eqnarray}
c_{2n}^2 &=& C ~+~ N^{-\nu}f_m(z) ~+~ N^{-2\nu}g_m(z) ~+~ 
N^{-3\nu}h_m(z) ~+~ \ldots \cr
c_{2n+1}^2 &=& \bar{C} ~+~ N^{-\nu}\bar{f}_m(z-N^{-\nu}) ~+~ 
N^{-2\nu}\bar{g}_m(z-N^{-\nu}) ~+~ N^{-3\nu}\bar{h}_m(z-N^{-\nu}) ~+~ \ldots 
\label{moorexp}
\end{eqnarray}
for $2n\approx N$, where the scaling variable $z$ is defined by
\beq
z ~\equiv~ N^{2m\nu}\left(1-\frac{2n}{N}\right)\label{zdef}
\eeq
with the exponent
\beq
\nu ~\equiv~ \frac{1}{2m+1}
\eeq
at $m$th multicriticality. Here $f_m(z),g_m(z),h_m(z)$ and the corresponding
barred quantities are assumed to be smooth functions. 

Consider first the case $m\!=\!1$. 
At the critical point, which is at $\frac{2n}{N}=1$, the two branches for the
even and odd recursion coefficients meet in the planar limit, which implies
$C=\bar{C}$.
Inserting eqs. (\ref{moorexp}) into the string equation (\r{stringexpl})
for even and odd $n$, and expanding consistently in $N^{-\nu}$ up to third 
order, yields the following set of equations:
\beq
C ~=~ \bar{C} ~=~ \frac{1}{4} ~,
\eeq
which is nothing else than the statement $2c_N\! =\! a +\ldots $. 
Furthermore we get
\beqn
\bar{f}_1(z) &=& -f_1(z) \\
g_1(z)+\bar{g}_1(z) &=& 2f_1(z)^2 -\frac{1}{8}z \l{ggb} \\
0 &=& f_1(z)\left( 16f_1(z)^2 - z\right) - 2f_1''(z) +\frac{\alpha}{2}~, 
\label{PII}
\eeqn
where the latter is a Painlev\'{e} II equation \cite{DSS} for $\alpha =0$. 
The dependence on 
$h_1(z)$ and $\bar{h}_1(z)$ has been eliminated using the equations from even
and odd $n$. Due to the following observation \cite{CM}
\beq
c^2_{2n+q} + c^2_{2n+q+1} ~=~ 2C ~+~ N^{-2\nu}
\left((-1)^qf'_m(z) + g_m(z)+\bar{g}_m(z)\right) ~+~ \ldots ~,\l{sumcs}
\eeq
where $2n+q\approx N$, we just have to determine the combination 
$g_1(z)+\bar{g}_1(z)$ 
and the function
$f_1(z)$ in order to get the quantities $u_{\pm}$ and $v$ from the last
subsection. So we just have to solve the generalized Painlev\'e II equation
(\ref{PII}).

\noi
Consider first the case $\alpha\!=\! 0$.
The solution to eq. (\ref{PII}) can then be shown to be unique \cite{DSS,CDM}
once the following physical boundary conditions are imposed. First, 
in the usual planar limit $z\! \to\! \infty$, we take\footnote{With $f_1(z)$
being a solution also $-f_1(z)$ is a solution. We fix the sign by noting
that for any finite $N$ the branch of the coefficients $\cn$ with odd $n$ 
is always larger than the one with even $n$ \cite{LE}.}
$f_1(z) = -\sqrt{z}/4$. This follows directly from solving the string equation
(\ref{stringexpl}) in the large-$N$ limit. 
Next, we demand that there are no singularities on the real $z$-axis.
(This is equivalent to enforcing $f_1(z) = 0$ at $z\!=\!-\infty$).
Because the boundary conditions in this manner are given at the opposite
ends of the interval (here the real line), and because the solution is
unstable towards the development of singularities once slightly improper
boundary conditions are imposed, it is not entirely trivial
to compute the correct solution numerically. We have found that the
discretized, iterative, procedure proposed in ref. \cite{BMP} for the
odd-order generalization of 
Painlev\'{e} I equation works well here too. In fig.~3
we show a plot
of the solution to eq. (\ref{PII}) found in this way. As a check, we 
display on the same figure a numerical determination of $f_1(z)$ directly
from the recursion coefficients $c_n$, evaluated to high order. 
For detailed definition of the parameters used to obtain this numerical 
solution, we refer to ref. \cite{BMP}.
\vspace{1cm}

\centerline{\epsfxsize=8cm 
\epsfbox{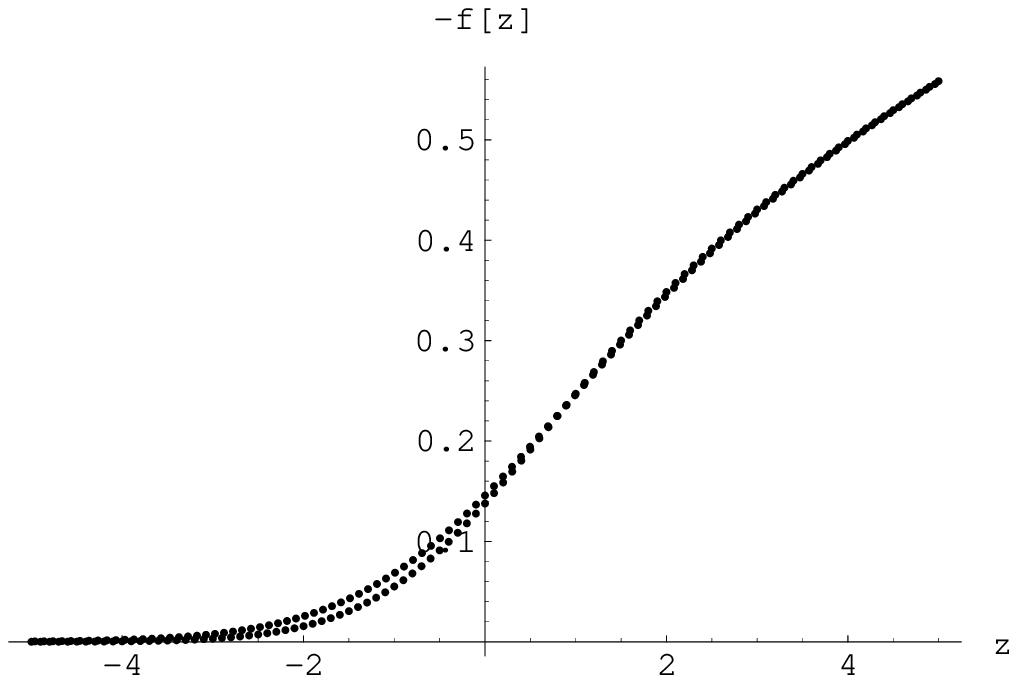}}
\noi
{\small Figure 3: Numerical solution to
Painlev\'{e} II equation $f(z) (16 f^2(z)-z)-2f''(z)=0$
(lower dots). We have used the parameters:
cutoff$=5$, mesh$=1/10$,  
calibration parameter $h=1/300$,  
number of iterations $\nu=400\sim 500$, 
and extrapolated the result to $\nu=\infty$.
Explicit values of $f_n = {N^{{\frac{1}{3}}}}\,
   \left(
  {2{{c^2_n}}{}}
- {{{c^2_{n-1}}}{}} 
- {{{c^2_{n+1}}}{}} \right)/4$
for the critical potential (\r{V1}) with $\alpha=0$, $N=1024$
are also plotted (upper dots). 
}

\noi
Inserting $g_2\!=\!-8$ and $g_4\!=\!16$ for the minimal $m\!=\!1$
multicritical potential, we get for $N$ even\footnote{For $N$ odd one has to 
replace $\pm\to\mp$ on the r.h.s. and $v\to -v$.}
\begin{eqnarray}
u_{\pm} &=& 16[2f_1(0)^2 \pm f_1'(0)] \cr
v &=& 32f_1(0) ~.
\end{eqnarray}
Note that the leading contributions (proportional to $N^{\frac{2}{3}}$
and $N^{\frac{1}{3}}$, respectively) to the above quantities vanish
at the critical point. The first corrections, from eq. (\r{sumcs}), are
precisely of the required power to render both $u_{\pm}$ and $v$
finite in the $N\! \to\! \infty$ limit. Yet higher order corrections
can indeed be ignored here.
If we substitute the solution to the Painlev\'{e} II equation, the full
differential equation (\ref{diffx}) for the orthogonal polynomials in the
microscopic limit is thus known in this case. Numerically we find from the 
above solution $u_+ \simeq -1.105 , u_- \simeq 2.211 $ and 
$v \simeq -4.207$. 
In fig.~4 we plot the solution to the equation
(\ref{diffx}) obtained in this way. On the same figure we also show the 
wave function for $\psi_N(x)$ 
constructed directly by Gramm-Schmidt orthonormalization.
The 
two functions are seen to agree to very high precision. This gives us complete
confidence in our procedure.\\

\noindent
\begin{minipage}{\textwidth}
\begin{center}
 \leavevmode
 \figinclude{237}{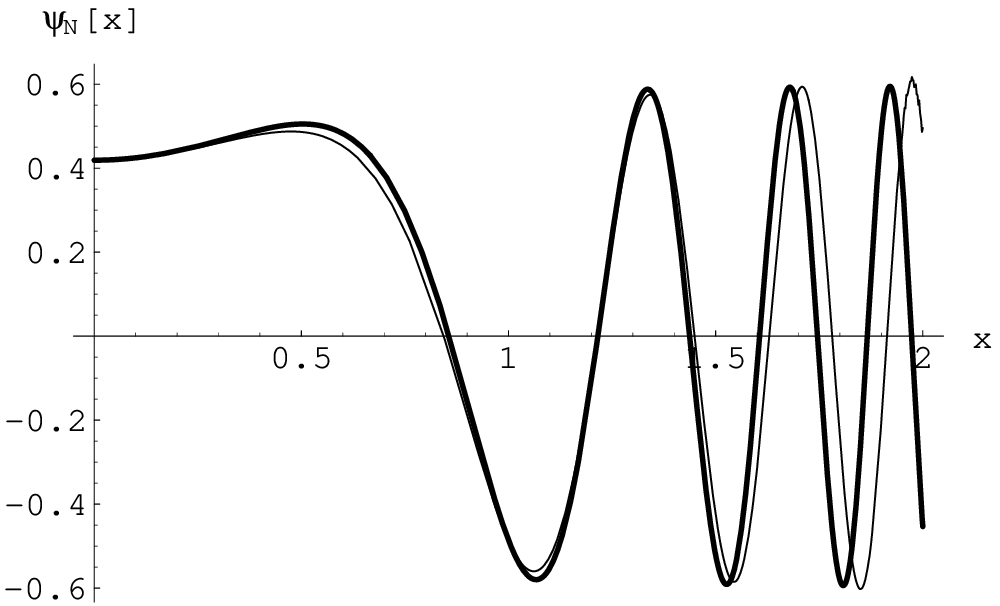}
 \figinclude{237}{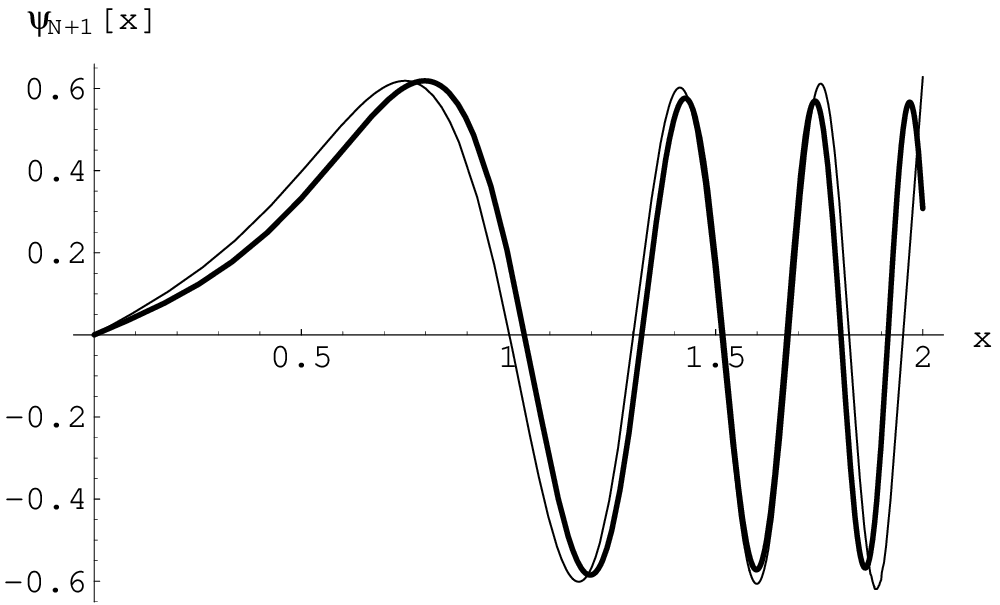}\\

\end{center}%\medskip
\noindent 
{\small 
Figure 4: Numerical solutions to the linear differential equation
(\ref{diffx}) using data from Painlev\'{e} II (bold lines).
Their vertical scales are chosen to optimize the matching with
the microscopic wave functions ($N=48$)
constructed from the critical potential with $\alpha=0$
by Gramm-Schmidt orthonormalization (real lines).
}
\end{minipage}

\noi
The microscopic correlation functions are constructed out
of these limiting functions for the wave functions.
In fig.~5 we exhibit a microscopic spectral density 
$\rho_S(x)\sim (\psi_N(x)\psi_{N-1}'(x)-\psi_{N-1}(x)\psi_N'(x))$:\\

\centerline{\epsfxsize=8cm \epsfbox{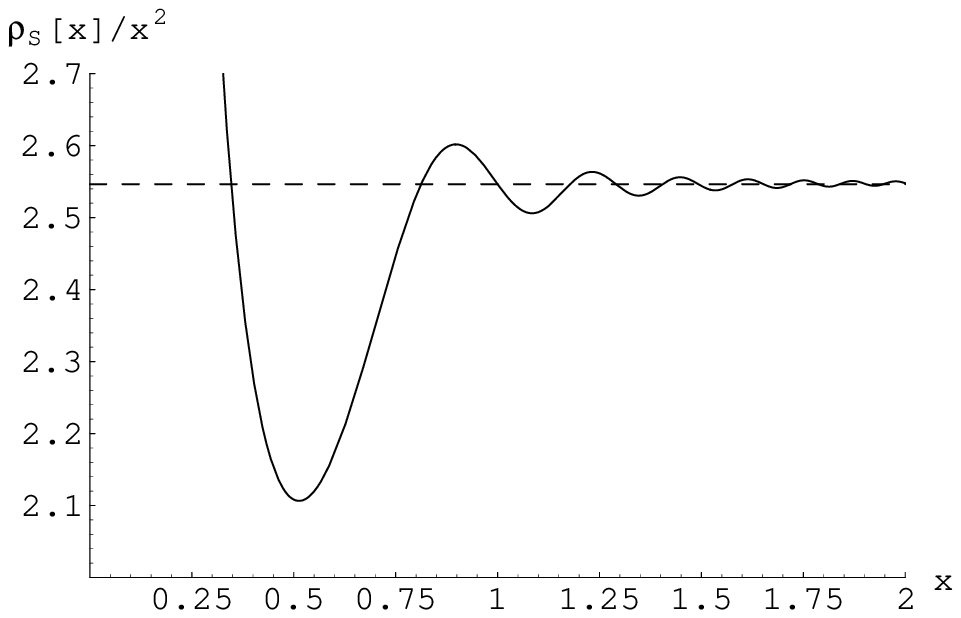}}
\noi{\small Figure 5:
Microscopic spectral density for the $m=1$ criticality, divided by $x^2$ 
in order to take out the trivial $x^2$-growth which is a consequence
of the matching-on to the macroscopic spectral density.
The normalization is determined by requiring
$\lim_{x\rightarrow\infty}\rho_S''(x)=\rho''(0)=16/\pi$
(dotted line).}

\noi
Next, consider $e.g.$, the case $\alpha\!=\! 1$. Appropriate boundary 
conditions for the generalized Painlev\'{e} II equation are easily established.
In the planar limit, here reached at $z\!\to\! \pm\infty$, the $\alpha$-term
does not contribute, as it is suppressed by $1/N$ compared with the rest
of the potential $V(\lambda)$. The boundary conditions are therefore
precisely the same as for $\alpha\!=\!0$:
\begin{eqnarray}
f_1(z) &\to& 0 ~~~~~~~{\mbox{\rm as}}~~ z \to -\infty ~,\label{PII1st}\\
f_1(z) &\to& -\sqrt{z}/4 ~~~~~~~{\mbox{\rm as}}~~ z \to \infty ~.\label{PII2nd}
\end{eqnarray}
For numerical purposes more refined boundary conditions are convenient:  
\begin{eqnarray}
f_1(z) &=& \alpha/(4z) ~~~~~~~{\mbox{\rm as}}~~ z \to -\infty ~ , 
\label{PII1stref}\\
f_1(z) &=& -\frac{1}{8}\left( \frac{z}{3y(z)}+y(z)\right) 
           \ +\ \frac{i\sqrt{3}}{8}\left( \frac{z}{3y(z)}-y(z)\right) 
~~~~~~~{\mbox{\rm as}}~~ z \to \infty ~, \nonumber\\
&&\mbox{with}~~~~~y(z)\equiv\left(-\alpha + \sqrt{{\alpha^2} - z^3/27} 
\right)^{\frac{1}{3}} 
\label{PII2ndref}
\end{eqnarray}
This last condition is found from solving the algebraic 3rd order equation
obtained from eq. (\ref{PII}), self-consistently setting
$f_1''(z) \sim 0$ as $z \to \infty$, and choosing the root which reduces
to $f_1(z) = -\sqrt{z}/4$ in the limit $z\!\to\! \infty$. In fact, because
the solution (\ref{PII2nd}), extended to the negative $z$-axis by 
\beq
f_1(z) ~=~ \frac{z}{12y(z)} + 
\frac{1}{4}y(z)
~~~~~~~{\mbox{\rm for}}~~ z < 0 ~,
\eeq
has $f_1''(z) \approx 0$ everywhere, and
automatically satisfies boundary condition (\ref{PII1st}), it
is almost exact. Numerically it is therefore an extremely efficient 
starting point for finding the full solution to eq. (\ref{PII}). 
It becomes exact in the limit $\alpha\!\to\!\infty$ 
(see also below).
  
\noi
In fig.~6 we plot the solution of the differential equation (\ref{diffx})
for $\alpha\!=\! 1$, and compare it with the solution
explicitly solving the recursions relations, using very 
for large values of $N$. We again
find perfect agreement. \vspace{1cm}

\centerline{\epsfxsize=8cm 
\epsfbox{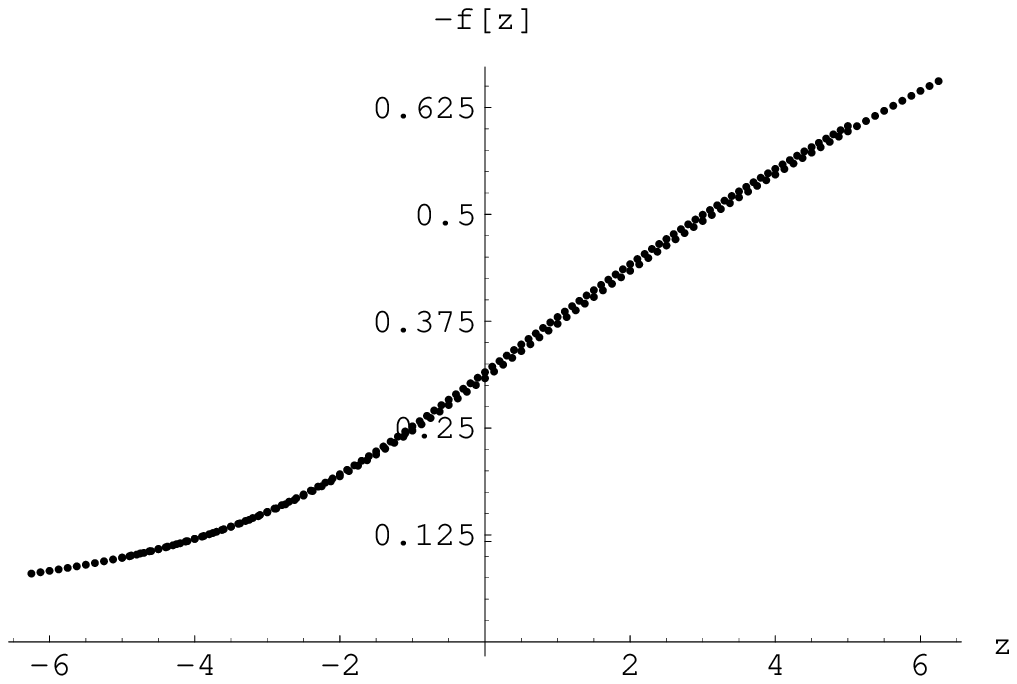}}
\noi
{\small Figure 6: Numerical solution to
Painlev\'{e} II equation $f(z) (16 f^2(z)-z)-2f''(z)+1/2=0$
(lower dots). We have used the same parameters as in fig.~3.
%cutoff$=5$, mesh$=1/10$,  
%calibration parameter $h=1/300$,  
%(see \r{BMP} for definition), and
%number of iterations $\nu=400\sim 500$, 
%and extrapolated the result to $\nu=\infty$.
Explicit values of $f_n$ 
for the critical potential with $\alpha=1$, $N=512$
are also plotted (upper dots). 
}

\noi
This is also a good point to check that the
$\alpha$-symmetry (\ref{asym}) is satisfied once we take the solution to
the generalized Painlev\'{e} II equation into account.  It follows from
eqs. (\ref{asym}) and (\ref{PII}) that the required relation is
(indicating the $\alpha$-dependence explicitly by a subscript):
\beq
f_{1(\alpha)}'(z) + 2f_{1(\alpha)}(z)^2 ~=~
-f_{1(\alpha-1)}'(z) + 2f_{1(\alpha-1)}(z)^2 ~.\label{asymf_1}
\eeq
This is obviously a highly non-trivial identity which the appropriate
solution to the generalized Painlev\'{e} II equation (\ref{PII}) must
satisfy. In the limit $\alpha
\!\to\! \infty$ the simple algebraic solution (\ref{PII2ndref}) approaches,
for fixed $z$, the full solution to the generalized Painlev\'{e} II equation
(\ref{PII}). It is straightforward to verify that the algebraic solution
(\ref{PII2ndref}) indeed satisfies the identity (\ref{asymf_1}), order
by order in a large-$\alpha$ expansion. Moreover, by performing
two Miura transformations we can actually prove that the relation 
(\ref{asymf_1}) is exact. The two transformations are
\begin{eqnarray}
M(z) &=& ~f_1'(z) + 2f_1(z)^2 \l{Miura1}\\
M(z) &=& -f_1'(z) + 2f_1(z)^2 ~~~and~~{\mbox{\rm shift}}~~ \alpha \to 
\alpha - 1 ~.\l{Miura2}
\end{eqnarray}
Under these two different transformations the generalized Painlev\'{e} II 
equation
({\ref{PII}) turns into the {\em same} equation:
\beq
\alpha(1-\alpha) + 2z^2M(z) - 32zM(z)^2 + 128M(z)^3 - 4M'(z)
+16\left[M'(z)\right]^2 + 4zM''(z) - 32 M(z)M''(z) ~=~ 0 ~.
\eeq
The two functions (\ref{Miura1}) and (\ref{Miura2}) thus satisfy the
same differential equation. The two boundary conditions are of course 
inherited from those of $f_1(z)$, and one can readily check from eqs.
(\ref{PII1st})-(\ref{PII2nd}) that the two functions
(\ref{Miura1}) and (\ref{Miura2}) also satisfy the same boundary
conditions at $z\!\to\! \pm\infty$. The two functions must therefore
be identical, and the relation (\ref{asymf_1}) has been proven. 

\noi
Note that, to the order at which we are working, the identity (\ref{asymf_1})
implies the following relation among the recursion coefficients (indicating
the $\alpha$-dependence explicitly by a superscript):
\beq
c^{(\alpha)2}_{2n} + c^{(\alpha)2}_{2n+1} ~=~ 
c^{(\alpha-1)2}_{2n+1} + c^{(\alpha-1)2}_{2n+2} ~.
\eeq
In this form one easily proves that the differential equation
for the wave functions (\ref{diffx}) respects the $\alpha$-symmetry
(\ref{asym}). 

\vspace{0.2cm}
\subsection{\sc Universality}

Up to now we have only been investigating the most simple example,
$m=1$ multicriticality with the minimal potential. In order to prove 
universality, one has to show that all perturbations to
the $m\!=\!1$ minimal potential that remain within the $m\!=\!1$ class
lead to the same universal differential equation (\ref{diffx}). 
It should be parameterized by only
one universal constant, $\rho^{(2m)}(0)$, in direct analogy with the 
noncritical case $m\!=\!0$ in \cite{us}. Since there is no longer a simple 
relationship as eq. (\ref{Arho}) at hand, this is a rather involved
task. Still, we have strong evidence that universality holds precisely in
the sense just described.
In order to illustrate this, we will give as an example a nonminimal
$m\!=\!1$ multicritical potential. We will also briefly outline the
situation for $m\!\geq\!2$ below.

\noi
When adding a sextic term to the minimal potential eq. (\ref{V1}),
\beq
V_1(\lambda) ~=~ \frac{1}{2}g_2\la^2 +\frac{1}{4}g_4\la^4 
+\frac{1}{6}g_6\la^6 ~, \label{V1n}
\eeq
the conditions for $m=1$ multicriticality and the corresponding
eigenvalue density read
\beqn
0 &=& g_2 +\frac{1}{2}g_4 +\frac{3}{8}g_6 ~, \nonumber\\
1 &=& \frac{1}{16}(g_4 + g_6) ~, \\
\rho(\la) &=& \frac{1}{2\pi}\left( (g_4+\frac{1}{2}g_6)\la^2+g_6\la^4 \right)
\sqrt{1-\la^2} ~.
\eeqn
The outcome of this modification is that the calculations of the previous
2 subsections exactly go through when replacing 
\beq
g_4 \ \rightarrow g_*\equiv g_4 + \frac{1}{2}g_6 ~,
\eeq
or, in other words, it depends only on one single parameter $\sim \rho''(0)$.

\noi
Without going through the details, where one can use the results for 
$m\!=\!2$ from appendix \ref{appB}, we define $u_{\pm}$ and $v$ as in eq. 
(\ref{uvm=2}), but with a scaling in powers of $N$ as in eq. (\ref{uv}). 
The final result reads
\beqn
u_{\pm} &=& g_* (2f^2(0) \pm f'(0)) ~, \nonumber\\
v\ &=& 2g_*f(0) ~, \label{uvnew}
\eeqn
where the generalized Painlev\'e II equation is now given by
\beq
0 ~=~ g_*f(z)^3 - zf(z) - g_*\frac{1}{8}f''(z) +\frac{\alpha}{2}~. \label{Pnew}
\eeq 
Using eq. (\ref{uvnew}) and $g_*$ instead of $g_4$ in eq. (\ref{Gasympt})
for $G_N$
and eq. (\ref{FG}) for $F_N$
provides the differential equation (\ref{diffx}) for
the $m=1$ nonminimal potential eq. (\ref{V1n}), where $g_*=\rho''(0)/\pi$.
Eqs. (\ref{uvnew}) and (\ref{Pnew}) and thus the differential
equation (\ref{diffx}) with (\ref{Gasympt}) will hold also for
all higher order perturbations of the minimal potential.

\noi
The essential ingredient in obtaining universality is thus the fact that
we still have the freedom of modifying the potential 
while remaining in the same universality class. The only quantity that
enters in the differential equation is, in the $m\!=\!1$ case, the second
derivative of the macroscopic spectral density at the origin, $\rho''(0)$.
All other details of the specific $m\!=\!1$ potential have disappeared.
By a simple rescaling of $\rho''(0)$, all $m\!=\!1$ critical potentials
should thus lead to the same universal functions.

\noi
The way this should generalize to the higher-$m$ case is as follows. An
alternative way of characterizing
multicritical potentials is by the scaling behavior
of the recursion coefficients $c_n^2$ near the ``critical point'',
where the two branches of the period-two solution merge. Here, in
the usual microscopic limit\cite{CM},
\beq
f_m(z) ~=~ -\frac{z^{1/2m}}{(2^{2m+1}(m+1))^{1/2m}} ~,~~~~~~~~~~~{\mbox{\rm
for}}~ z \to \infty \label{fmlimit}
\eeq
which is the generalization to higher $m$ of the behavior 
$\lim_{z\to\infty}f_1(z) \sim -\sqrt{z}/4$ 
we found explicitly in the $m\!=\! 1$ case. This scaling of
the recursion coefficients near the point where they merge, cf. eqs.
(\ref{moorexp}) and (\ref{fmlimit}), 
\beq
c_n^2 -C ~\sim~ \pm {\mbox{\rm const.}}~
\left(1 - \frac{n}{N}\right)^{\frac{1}{2m}}\label{cscaling}
\eeq
is {\em universal} in the sense that its power-law scaling 
depends only on the class $m$
of multicriticality (the positive solution corresponds to odd $n$, while
the negative one corresponds to even $n$). 
The coefficients $C\!=\!\bar{C}$ of eq. (\ref{moorexp}) will change with
the chosen $m$th multicritical potential, and so will the particular
normalization of (\ref{fmlimit}), but the scaling (\ref{cscaling})
will remain. The non-universal coefficients $C\!=\!\bar{C}$ drop out
of the differential equation for the orthogonal polynomials
simply by the requirement of $m$th order multicriticality. The next term
of the expansion (\ref{sumcs}) is what will enter in the differential
equation. Our statements about universality above translate into 
saying that for all $m$ this particular combination of solutions to the
generalized Painlev\'{e} II equations for $f_m(z)$ (and higher
functions of the expansion (\ref{moorexp})) will be proportional to
$\rho^{(2m)}(0)$. We have not proven this.

\vspace{0.2cm}
\subsection{\sc A Double-Scaling Limit}

As we have shown above, apparently non-leading terms from the coefficients
$c_n$ turn out to contribute in the multicritical microscopic limit. At
the core of this phenomenon stands the breakdown in this limit of the 
simple relation (\ref{Arho}) between the function $A_n$ (which enters 
directly into the differential equation for the polynomials), and the 
macroscopic spectral density $\rho(\lambda)$. The way this relation
breaks down at multicriticality makes it clear that the phenomenon is
more general, and directly related to the very definition of multicriticality.
So far we have used the most naive definition, where we first tune the 
couplings of the potential $V(\lambda)$ to their values at the $N\!=\!\infty$
multicriticality, and {\em then} take the $N \to \infty$ limit in the
differential equation for the polynomials. What we implicitly have shown
above is that the final result will depend on this precise procedure.
This should come as no surprise, since we in contrast to the
conventional case of a finite $\rho(0)$ really have to approach a
critical point at some particular critical coupling(s). The whole analysis
is then quite analogous to finite-size scaling in statistical
mechanics \cite{B}. Finite $N$ here plays the r\^{o}le of finite volume $V$,
and as in statistical mechanics the notion of a (multi)critical point
at finite $N$ is ambiguous. One can define it to be precisely at the
position of the infinite-$N$ (multi)critical point, or one can choose other
definitions (at the peak of the specific heat, or otherwise) that
all agree in the $N\!=\!\infty$ limit.

\noi
In order to remedy this situation, we clearly have to consider more
general ways of reaching multicriticality. Instead of starting out
with the couplings fixed precisely at the values corresponding to 
$N\!=\!\infty$ multicriticality, we can consider a continuum
of different schemes in which the couplings are tuned towards the
multicritical values at just such a rate as to contribute non-trivial
terms to the differential equation in the microscopic $N\to\infty$ limit.
Because $m$ couplings must be tuned to reach $m$th-order 
multicriticality, this program is however not very practical from
an analytical point of view. A much more economical procedure is to tune
only one parameter, which in a generic manner takes us away from and
toward the multicritical points. Following the conventions of matrix
models applied to 2-d quantum gravity, we shall call this
a double-scaling limit. It is, in a very precise sense, equivalent to
finite-size scaling theory in statistical mechanics \cite{DH}.  

\noi
Suppose, then, that we have chosen our potential $V(\lambda)$ to be
multicritical according to eq. (\ref{vprime}) for some given $M$. 
Consider now the partition function defined by
\beq
{\cal Z} ~=~ \int\!dM {\det}^{2\alpha}M ~e^{-(N/g) Tr V(M)} ~.
\label{Zg}
\eeq
When $g\!=\! 1$ this coincides with our previous definition, eq. (\ref{Z}).
We can tune $g$ towards $g_c\!=\! 1$ in such a way that
\beq
y ~\equiv~ N^{2m\nu}(g - 1)
\eeq
is kept fixed. This defines our double-scaling limit. The previous analysis
was simply restricted to $y\!=\! 0$. Now all solutions will be parameterized
by (functions of) this number $y$, which is kept fixed in the large-$N$ limit. 
The appropriate scaling variable is then
\beq
z ~\equiv~ N^{2m\nu}\left(1-\frac{2n}{N}g\right)\label{zgdef}
\eeq
instead of eq. (\ref{zdef}). This is completely analogous to the treatment
of multicritical limits at the edge of the spectrum by Bowick and 
Brezin \cite{BB}. We can now probe a double-scaling region where $g$ 
differs from unity only by an amount of the order $N^{-2m\nu}$, while
the eigenvalues $\la$ differ from zero by an amount of the order
$N^{-\nu}$. 
Essentially all of the previous discussion goes through unchanged in this
more general setting, with just some minor redefinitions. It follows
from the above that the sole effect of including a non-vanishing $y$ is to 
shift the scaling variable by $z \to z+y$. Accordingly, all of the
previous discussion goes through unchanged in this
more general setting, except that we should substitute the numerical values
of the Painlev'{e} II transcendent at $z=y$ instead of $z=0$ into the linear
differential equation (\ref{diffx}).

\vspace{0.2cm}
\subsection{\sc A multicritical mesoscopic range}\label{meso}

It is of some interest to compare the correct solution for the wave functions
$\psi_N(x)$ with what we would have obtained if we had ignored the
apparently subleading terms in the defining differential equation
(\ref{diffx}). Consider again for simplicity the case $m\!=\! 1$ and
$\alpha\!=\! 0$, for which we can compare the two solutions visually
(see fig.2). It is clear that although the two solutions differ 
in detail, especially very close to the origin, their qualitative features 
are nevertheless quite similar. This has a simple algebraic explanation.
The coefficient of the leading power of $A_N(\la)$ is proportional to $k(m)$,
which can be seen from eq. (\ref{vprime}) and the definition (\ref{An}).
Already for $m=1,2$ $k(m)$ is large compared to one, a fact which is improving 
for growing $m$:
\beq
1 \ \ll \ k(m) \ \longrightarrow\ 4\sqrt{2\pi} m^{\frac{3}{2}}~~~~~
\mbox{for}~~~~~~ m\to\infty \ .
\label{klim}
\eeq
If now the leading power of $A_N(\la)$ is dominating all the lower terms,
we will get back the relation (\ref{Arho}) to the eigenvalue density
as in the noncritical case and will therefore recover the ``naive''
differential equation from (\ref{Gnaiv}). The appropriate condition
for $x$ in this mesoscopic limit is
\beq
1 \ \ll \ k(m)~x^{2m} \ , \label{mesocond}
\eeq
with the simplified differential equation reading
\beq
\psi_N''(x) - \frac{2m}{x}\psi_N'(x) + 
\left(c_N^2k(m)x^{4m} + \frac{(-1)^N\al (2m+1)-\al^2}{x^2}\right) 
\psi_N(x) ~=~ 0 ~. \label{diffmes}
\eeq
This mesoscopic differential equation can be solved analytically
in terms of Bessel functions. Note that since it is an {\em approximate} 
equation it is no longer $\al$-symmetric. At growing multicriticality $m$ it
will approximate the true differential equation (\ref{diffex})
for smaller and smaller values of $x$, due to the behavior of
$k(m)$ eq. (\ref{klim}).

\noi
As it has been mentioned eq. (\ref{diffmes}) can be mapped to the following 
equation of Bessel type
\beq
\left( X^{-\beta}\frac{d}{dX} X^{\beta}\frac{d}{dX} ~+~1\right) 
X^{-\frac{\beta}{2}} \psi_N(X) ~=0~,
\eeq
where
\beq
X~=~ \frac{k(m)}{2(2m+1)}x^{2m+1}
=\frac{\pi \rho^{(2m)}(0)}{(2m+1)!} {x^{2m+1}}
 ~ ,
~~~~~~~ \beta ~=~ \frac{2(-1)^N\al}{2m+1} ~.
\label{Xdef}
\eeq
Boundary conditions are however not easily established.
The solution satisfying the usual boundary conditions, namely 
regularity at the origin and normalizability, reads
\beq
\psi_N(x) ~\sim~ 
\sqrt{X}\,
J_{\frac{\al}{2m+1}-\frac{(-1)^N}{2}}(X) ~,
\label{psimeso}
\eeq
which still has to be normalized. (Since the differential equation cannot
be trusted for small values of $x$, it is not obvious that the above
boundary conditions are suitable; we shall for illustration stick to 
them here.) The approximate kernel and the approximate spectral density 
are then given by
\beqn
K_{meso}(x,x')
&=&\lim_{N\to\infty}
{N^{-\frac{1}{2m+1}}}
K_N\left({N^{-\frac{1}{2m+1}}}x,
{N^{-\frac{1}{2m+1}}}x'\right)  \\
&=& 
\frac{\pi\rho^{(2m)}(0)}{2(2m+1)!}
\frac{(xx')^{m+\frac{1}{2}}}{(x-x')}
\left( 
J_{\frac{\al}{2m+1}+\frac{1}{2}}(X)
J_{\frac{\al}{2m+1}-\frac{1}{2}}(X')
~-~
J_{\frac{\al}{2m+1}-\frac{1}{2}}(X)
J_{\frac{\al}{2m+1}+\frac{1}{2}}(X')
\right) ~,\nonumber
\eeqn
and
\newpage
\beqn
&&\rho_{meso}(x) = 
\left(\frac{\pi\rho^{(2m)}(0)}{2(2m)!}\right)^2
\frac{x^{4m+1}}{2m+1} \times 
\label{finalrho}\\
&&~~~~~~~~
\left(J_{\frac{\al}{2m+1}+\frac{1}{2}}(X)^2 +
J_{\frac{\al}{2m+1}-\frac{1}{2}}(X)^2 -
J_{\frac{\al}{2m+1}+\frac{1}{2}}(X)J_{\frac{\al}{2m+1}-\frac{3}{2}}(X) -
J_{\frac{\al}{2m+1}-\frac{1}{2}}(X)J_{\frac{\al}{2m+1}+\frac{3}{2}}(X)
\right)~. 
\nonumber
\eeqn
Here the normalization has been fixed by requiring the matching condition
\beq
\lim_{x\to\infty}\rho_{meso}^{(2m)}(x)
~=~ \rho^{(2m)}(0) ~=~ (2m)!\frac{k(m)}{2\pi}~.
\eeq

\noi
The above ``mesoscopic'' spectral density of course has only
approximate validity. 
For $\alpha\!=\! 0$ it simply coincides with the 
macroscopic spectral density near the origin (a feature it shares with the
usual non-critical microscopic spectral density (\ref{first})), and is in this
sense ``exact''. 
For $\alpha/(2m+1)$ non-zero, but small,
it does define an intermediate range where the set in of the universal 
mesoscopic oscillations is correctly encoded, and the macroscopic spectral 
density still not reached. To give an example of the approximation achieved,
we show below, for $\alpha\!=\! 1$ and $m\!=\! 1$, 
the full microscopic spectral density (fig.~7) and the approximate
mesoscopic spectral density (fig.~8). The full microscopic spectral
density has been computed using the method described above, where, from the
solution to the Painlev\'{e} II equation, we find $u_+\!=\! 2.141$,
$u_-\!=\! 4.218$ and $v\!=\! -10.087$.

\centerline{\epsfxsize=8cm \epsfbox{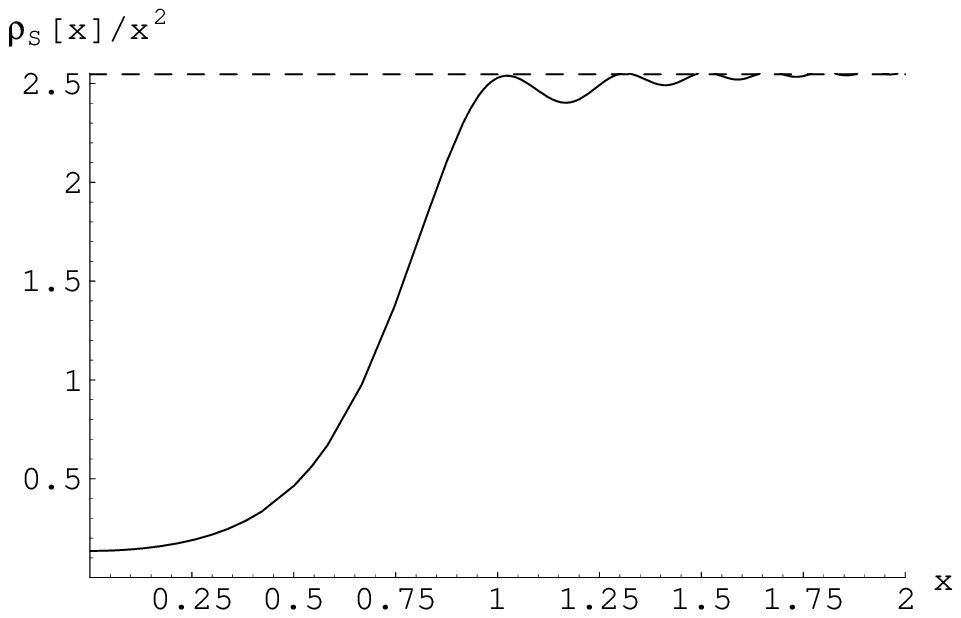}}
\centerline{{\small Figure 7:
Microscopic spectral density for
$m\!=\! 1$ criticality with $\alpha\!=\! 1$.}}

\centerline{\epsfxsize=8cm \epsfbox{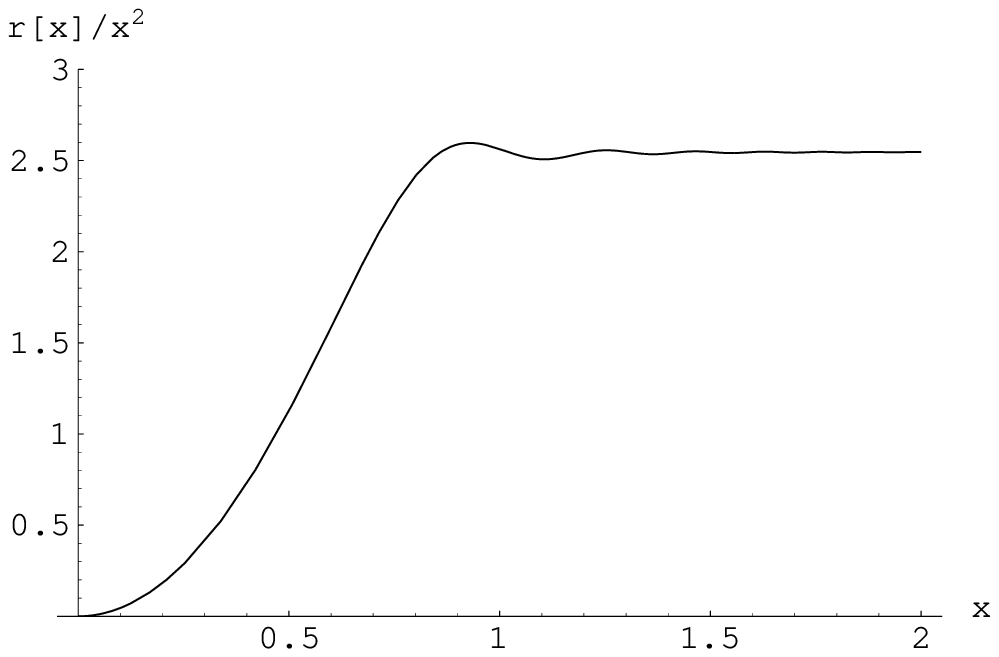}}
\centerline{{\small Figure 8: The approximate mesoscopic spectral density 
$\rho_{meso}(x)/x^2$ for
$m\!=\! 1$, $\alpha\!=\! 1$.}}

\noi
For $m=0$ ($k(0)=4$) the mesoscopic density 
reduces to the exact, noncritical {\it microscopic} 
density \cite{us}.

\setcounter{equation}{0}
\section{Multicritical Chiral Unitary Ensembles}

Once the above results have been established, it is only a small step to
extend them from the unitary ensemble to the chiral unitary ensemble. This is
due to the fact that the partition functions, and their associated orthogonal
polynomials, for the two different ensembles can be related to each other 
through appropriate shifts of the arguments. Let us recall \cite{V} that the 
pertinent matrix integral for theories in even dimensions, gauge groups 
$SU(N\!\geq\!3)$, and fermions in the fundamental representations of these 
gauge groups, is described by complex matrices \cite{TRM}:
\beq
{\cal Z} ~=~ \int\! dM {\det}^{N_f}M~ 
\exp\left[-\frac{N}{2} Tr \tilde{V}(M^2)\right] ~,
\label{cZ}
\eeq
with an even potential
\beq
\tilde{V}(M^2) = \sum \frac{\tilde{g_k}}{k} M^{2k} ~.
\eeq
Here $M$ is a $(2N\! +\! |\nu |)\!\times\! (2N\! +\! |\nu |)$ 
block hermitian matrix of the form
\beq
M ~=~ \left( \begin{array}{cc}
              0 & W^{\dagger} \\
              W & 0
              \end{array}
      \right) 
\eeq
where $W$ itself is a rectangular complex matrix of size 
$N\times(N\! +\! |\nu|)$.
The translation of parameters into gauge theory language is that $\nu$
represents the topological charge, and the space-time volume equals
$2N\! +\! |\nu|$, the size of the matrix $M$. The 
integration measure in eq. (\ref{cZ}) is the Haar measure of $W$. This
measure and the integrand of eq. (\ref{cZ}) both are invariant under
left and right multiplications:
\beq
W \to TWU~, ~~~~~~~~ U \in U(N)~, ~~~~ T \in U(N\! +\! |\nu|) ~.
\eeq

\noi
It is convenient to rewrite the matrix integral (\ref{cZ}) in terms of
the eigenvalues 
of the hermitian matrix $H \equiv W^{\dagger}W$ \cite{TRM}:
\beqn
{\cal Z} &=& \int\! dWdW^{\dagger}  {\det}^{N_f}(W^{\dagger}W) ~ 
\exp\left[-N Tr \tilde{V}(W^{\dagger}W)\right] ~, \nonumber\\
&\sim& \int_0^{\infty}\! \prod_{i=1}^N \left(d\lambda_i
\lambda_i^{\alpha}e^{-N\tilde{V}(\lambda_i)}\right)\left|{\det}_{ij}
\lambda_j^{i-1}\right|^2 ~, \label{Zpos}\\
&\sim& \int_{-\infty}^{\infty}\! \prod_{i=1}^N \left(dz_i
|z_i|^{2\alpha +1}e^{-N\tilde{V}(z_i^2)}\right)\left|{\det}_{ij}
z_j^{2i-2}\right|^2 ~, \label{Zreal}
\eeqn
where $\alpha \equiv N_f\! +\! |\nu|$. Here the fact has been used, that one
can switch from real positive eigenvalues $\lambda_i$ 
to real eigenvalues $z_i$ \cite{TRM}.
In the latter form, the matrix integral
can be analyzed directly in terms of the unitary ensemble considered
in the previous section. 

\noi
Let us introduce a set of even polynomials 
$\tilde{P}_l^{(\alpha)}(z^2)$ 
orthonormal with respect to the measure
$d\omega (z)\equiv dz|z|^{2\alpha +1}\exp(-NV(z^2))$ on the real line.
Then we can replace the Vandermonde determinant as usual,
but this time by a determinant of only even polynomials 
$\tilde{P}_l^{(\alpha)}(z^2)$. 
In doing so we can directly use the even 
subset\footnote{We will also encounter the odd polynomials when
calculating the kernel via the Christoffel-Darboux formula.}
of the orthonormal polynomials eq. (\ref{ONP}), which we have already
calculated, by shifting $\alpha \rightarrow \alpha + \frac{1}{2}$:
\beq
\tilde{P}_l^{(\alpha)}(z^2) ~\equiv~ P_{2l}^{(\alpha+\frac{1}{2})}(z) ~.
\label{idOP}
\eeq
In going from eigenvalues $\lambda$ to $z$ in eq. (\ref{Zreal})
we have an even potential
$V(z^2)$ as we should, in contrast to the real positive picture
(\ref{Zpos}). In order to justify the identification (\ref{idOP})
also in the multicritical case we still need a relation
between the respective critical potentials. It is given by \cite{DJM}
\beq
\tilde{V}_m(\la^2) ~=~ 2 V_m(\la) \ ,
\label{idVs}
\eeq
The critical potential for the chiral unitary
ensemble has been calculated by imposing 
\beq
\rho_m(z) ~=~ \frac{1}{4\pi} k(m) z^{2m} \sqrt{1-z^2} ~.
\eeq
for the eigenvalue density on R, exactly as in the unitary case eq. 
(\ref{vprime}). The factor of two in eq. (\ref{idVs}) 
can be traced back to the difference in the saddle point
equations of the two ensembles \cite{DJM}. 
Consequently the wavefunctions of the chiral unitary ensemble 
at $m$-th multicriticality can be taken
from the unitary ones eq. (\ref{defpsi}) simply by shifting
$\alpha \rightarrow \alpha + \frac{1}{2}$:
\beqn
\tilde{\psi}_l^{(\alpha)}(z^2) &\equiv& |z|^{\alpha +\frac{1}{2}} 
e^{-\frac{N}{2}\tilde{V}_m(z^2)}\tilde{P}_l^{(\alpha)}(z^2) \nonumber\\
&=& |z|^{\alpha +\frac{1}{2}}
e^{-NV_m(z)}P_{2l}^{(\alpha+\frac{1}{2})}(z) \nonumber\\
&=& \psi_{2l}^{(\alpha+\frac{1}{2})}(z) ~~,\label{psich}
\eeqn
where we have identified the couplings $\tilde{g}_k \equiv g_k$
according to eq. (\ref{idVs}). The wavefunctions of the unitary ensemble 
$\psi_{2l}^{(\alpha+\frac{1}{2})}(z)$ 
now have to be evaluated at
$2N$ instead of $N$. They are determined by the corresponding universal
differential equation as being discussed in great detail in the previous
section.

\noi
The eigenvalue density in the microscopic limit is obtained from the kernel
expressed in terms of even {\it and} odd
wave functions $\psi_{k}^{(\alpha+\frac{1}{2})}(z)$: 
\beqn
K_N(z^2,w^2) &\equiv& |zw|^{\alpha +\frac{1}{2}}
e^{-\frac{1}{2}N(\tilde{V}_m(z^2)+\tilde{V}_m(w^2))}\sum_{l=0}^{N-1}
\tilde{P}_l^{(\alpha)}(z^2)\tilde{P}_l^{(\alpha)}(w^2) \nonumber\\
&=& \sum_{l=0}^{N-1}\psi_{2l}^{(\alpha+\frac{1}{2})}(z)
\psi_{2l}^{(\alpha+\frac{1}{2})}(w) \nonumber\\
&=&c_{2N} c_{2N-1} \frac{1}{z^2-w^2}
\left( \psi_{2N}^{(\alpha+\frac{1}{2})}(z)
\psi_{2N-2}^{(\alpha+\frac{1}{2})}(w) 
~-~ \psi_{2N-2}^{(\alpha+\frac{1}{2})}(z) 
\psi_{2N}^{(\alpha+\frac{1}{2})}(w)\right) \nonumber\\
&=&c_{2N}\frac{1}{z^2-w^2}
\left( \psi_{2N}^{(\alpha+\frac{1}{2})}(z)
\ w\ \psi_{2N-1}^{(\alpha+\frac{1}{2})}(w) 
~-~ \ z\ \psi_{2N-1}^{(\alpha+\frac{1}{2})}(z) 
 \psi_{2N}^{(\alpha+\frac{1}{2})}(w)\right) ~. \label{kerch}
\eeqn
Here we have made use of the Christoffel-Darboux identity and the recursion 
relation for the wave functions. The advantage of the last expression
is that we have avoided to expand $\psi_{2N-2}^{(\alpha+\frac{1}{2})}(x)$ in 
$1/N$. Imposing the same scaling for the real eigenvalues $z$ as in 
eq. (\ref{mresc}),
\beq
x ~=~ N^{\frac{1}{2m+1}}z 
\eeq
we obtain
\beqn
\rho_s(x) &=& \lim_{y\to x}\lim_{N\to\infty}N^{-\frac{1}{2m+1}}
K_N\left(x^2N^{-\frac{2}{2m+1}},y^2N^{-\frac{2}{2m+1}}\right) \\
&=& \lim_{N\to\infty}\frac{c_{2N}}{2} 
\left( \psi_{2N}^{(\alpha+\frac{1}{2})}(x)'
\psi_{2N-1}^{(\alpha+\frac{1}{2})}(x) ~-~ \psi_{2N}^{(\alpha+\frac{1}{2})}(x)
\psi_{2N-1}^{(\alpha+\frac{1}{2})}(x)' ~-~
\frac{1}{x}\psi_{2N}^{(\alpha+\frac{1}{2})}(x)
\psi_{2N-1}^{(\alpha+\frac{1}{2})} \right) \nonumber
\label{rhosch} ~.
\eeqn

\vspace{0.2cm}
\subsection{\sc The multicritical mesoscopic range}

Since in the unitary case we could obtain analytic expressions
for the wave functions in an intermediate, mesoscopic range we can do
the same here, using the results of subsection \ref{meso}.
The range of applicability is thus given again by eq. (\ref{mesocond}).
The unnormalized, analytic expressions for the mesoscopic wave functions
eq. (\ref{psimeso}), which are needed here read (see our discussion about
boundary conditions in the previous section):
\beqn
\psi_{2N}^{(\alpha+\frac{1}{2})}(x) &\sim& \sqrt{X}
J_{\beta-\frac{1}{2}}(X) \ , \nonumber\\
\psi_{2N-1}^{(\alpha+\frac{1}{2})}(x) &\sim&\sqrt{X}
J_{\beta+\frac{1}{2}}(X) \ ,
\eeqn
where
\beq
X \equiv \frac{k(m)}{2(2m+1)}x^{2m+1}
=
2\pi {{\rho }^{(2m)}}(0){\frac{{x^{2m+1}}}{\left( 2m+1 \right) !}}
~~ ,~~~~~~~~~~~ 
\beta\equiv\frac{2\alpha+1}{2(2m+1)}~.
\eeq
Inserting these expressions into eqs. (\ref{kerch}) and (\ref{rhosch}) 
we obtain the following expressions for the approximate mesoscopic
kernel and spectral density 
\beqn
K_{meso}(x,x')&\equiv&
\lim_{N\to\infty}N^{-\frac{1}{2m+1}}
K_N\left(x^2N^{-\frac{2}{2m+1}},x'^2N^{-\frac{2}{2m+1}}\right)
\nonumber\\
&=& \frac{2\pi\rho^{(2m)}(0)}{(2m+1)!}
\frac{(xx')^{m+\frac{1}{2}}}{x^2-x'^2}
\left(x\ J_{\beta-\frac{1}{2}}(X')
J_{\beta+\frac{1}{2}}(X) ~-~
x'\ J_{\beta-\frac{1}{2}}(X)
J_{\beta+\frac{1}{2}}(X')\right)
\eeqn
and
\beq
\rho_{meso}(x) =
{\frac{\pi {{\rho }^{(2m)}}(0){x^{2m}}}{2\left( 2m \right) !}}
\left(
{X}\left( 
{{{J_{\beta+{\frac{1}{2}}}}(X)}^2} + 
{{{J_{{\beta-\frac{1}{2}}}}(X)}^2} \right)
-{\frac{2\alpha }{2m+1}}
{J_{\beta+{\frac{1}{2}}}}(X)
{J_{\beta-{\frac{1}{2}}}}(X)
\right) ~.
\label{finalrhoch}
\eeq
The normalization has been fixed by requiring again
\beq
\lim_{x\to\infty}\rho_{meso}^{(2m)}(x) ~=~ \rho^{(2m)}(0)
~=~ (2m)!\frac{k(m)}{4\pi}~.
\eeq
For $m=0$ the above mesoscopic density (\ref{finalrhoch}) 
reduces to the exact noncritical {\it microscopic} density \cite{us}.

\noi
In fig.\ 9 we exhibit the exact microscopic density (\ref{rhosch})
for $m=1$ and $\alpha=0$,
constructed by solving the Painlev\'e II equation for
$\alpha=1/2$, and
the mesoscopic approximation to it eq. (\ref{finalrhoch}).\vspace{5mm}

\centerline{\epsfxsize=8cm \epsfbox{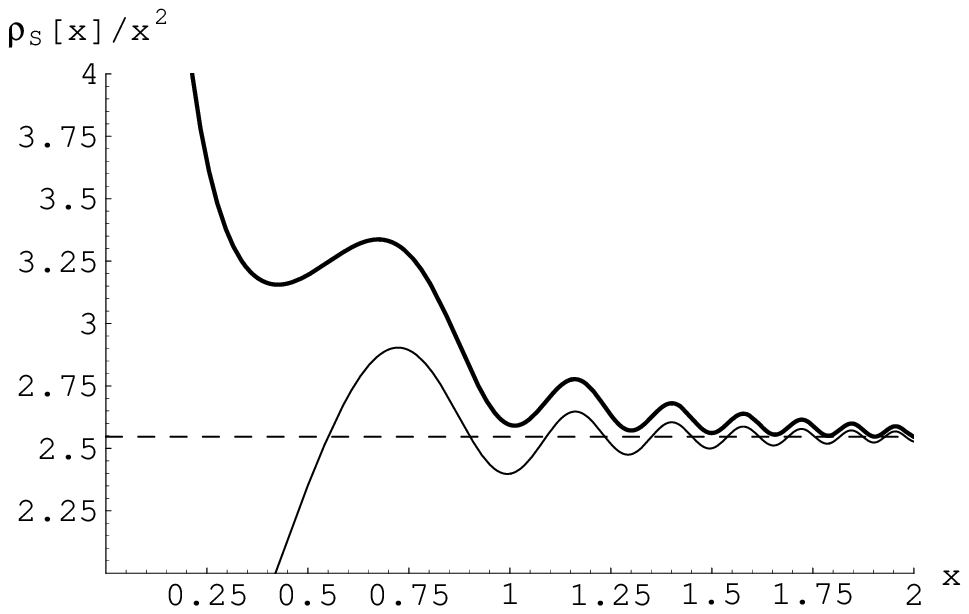}}
\noi{\small Figure 9:
Microscopic spectral density for the $m=1$ critical
chiral unitary ensemble with $\alpha=0$ (bold line)
and its mesoscopic approximation (real line).
The normalization is determined by requiring
$\lim_{x\rightarrow\infty}\rho_S''(x)=\rho''(0)=8/\pi$
(dotted line).
}

\setcounter{equation}{0}
\section{Conclusions}

In this paper we have investigated random matrix models which may be related
to QCD-like gauge theories with unbroken chiral or flavor symmetries.
{}From the Banks-Casher relation \cite{BC} the vanishing of the corresponding 
order parameter $\langle \bar{\psi} \psi \rangle$ implies a 
vanishing spectral density of the Dirac operator at the origin.
The case where this density  vanishes as an even power can be mapped 
to the multicritical points in the appropriate matrix models if one 
follows the same arguments as in \cite{V} for the broken phase. 

\noi
The new universal results we find for multicritical
matrix models in the microscopic large-$N$ limit may suggest a
universal behavior for the microscopic density of the Dirac operator itself.
In contrast to the case of broken symmetries, there are here no known
spectral sum rules of these theories with which to compare the results
from random matrix theory. The relevance of these results for the
Dirac spectrum of theories with unbroken chiral/flavor symmetries
is therefore still far from obvious, and we can at present only view
it as an intriguing conjecture. It would be particularly interesting to
investigate if an approach analogous to that of a chiral Lagrangian
\cite{LS} exists in this case. 

\noi
The detailed universality predictions,
the listing of universality classes, and the ``critical exponents''
$\nu = 1/(2m+1)$ can all be tested in lattice gauge theory with an
appropriate fermion prescription. 
In contrast to the noncritical case \cite{us}, the universal results
for the microscopic spectral densities
are no longer given in terms of elementary functions. Instead, we derive 
universal differential equations for the orthogonal polynomials of the
multicritical models, which then yield the spectral density from its kernel. 
For the practical purpose of comparison with lattice gauge theory data,
the generalization of the present formulation to that of massive fermions
may be highly advantageous. Recently the massive case for non-critical
potentials was worked out in detail for the chiral unitary ensemble
and proven to be universal \cite{DN} in the appropriate scaling limit
(see also ref. \cite{WGW} for an explicit computation in the Gaussian case). 
The same universality
of the massive case extends also to unitary ensemble \cite{DN1}. The
multicritical cases for both ensembles with non-zero masses can be worked 
out completely analogously using the wave functions derived in this paper.
The universality arguments also generalize accordingly.

\noi
We have discussed here in great detail the solution 
for first order multicriticality with an arbitrary number of massless fermions,
giving evidence for the universality of the determining differential 
equation and thus for the microscopic spectral density. 
For higher order criticality we have derived the general form of 
the differential equation and conjectured its universality. With
growing multicriticality $m$ the analysis becomes increasingly involved,
but all the necessary ingredients for carrying the computations through
for arbitrary $m$ have been presented here.
Moreover, in a certain mesoscopic large-$N$ limit we have been able to 
give approximate, analytic expressions for the universality classes of 
the orthogonal polynomials as well as the microscopic density 
for any order of multicriticality and any number of fermions.

\noi
In the noncritical case, the relevant scale that determines the microscopic
limit is the spectral density at the origin, $\rho(0)$. All eigenvalues are
in that case blown up near $\la\!=\! 0$ by the scale $\rho(0)N$. 
As could have been guessed
from the outset, in the $m$th multicritical case the relevant scale is
set by the first non-vanishing derivative of the macroscopic spectral
density at the origin, $\rho^{(2m)}(0)$. All eigenvalues must be blown
up by a scale determined by this number. In addition, and this is what
requires a more detailed analysis, we have seen that the eigenvalues must 
be blown up with the volume $N$ raised to the exponent $\nu$.
The noncritical case is simply that of $m\!=\! 0$. Viewed from this 
perspective,
the universality classes derived in this paper are the completely
natural generalizations of those derived previously \cite{V,V1}. Universality
simply translates into the freedom of redefining only one scale, 
$\rho^{(2m)}(0)$, in order to leave all results invariant.

\noi
One might think that the most obvious application of the universal results
described in this paper would be in connection with chiral symmetry
restoration at finite temperature or density. Because one parameter
needs to be tuned in each case (temperature $T$, or chemical potential
$\mu$), one could believe that these two cases, when occurring at continuous
phase transitions, would be good candidates for the $m\!=\!1$ universality 
class. We are mildly skeptical about this, and have instead, in the 
introduction, suggested what seem to us to be more
promising candidates. There has already been a large amount of work on
possible extensions of the by now essentially established $T\!=\!0$
results to the $T\neq 0$ and $\mu\neq 0$ situations (see, $e.g.$, ref.
\cite{Tneq0} for just a limited selection of papers, and also others cited in
ref. \cite{Rev}). 

\vspace{1cm}

\noi
{\large \bf{Acknowledgments}}

\noi
We thank T.R. Morris for discussions at an early stage of this work.
The work of G.A. is supported by European Community grant no.
{\small ERBFMBICT960997}. 
Furthermore he wishes to thank the Niels Bohr Institute
for its warm hospitality during several stays.
The work of S.M.N. is supported in part by 
JSPS Postdoctoral Fellowships for Research Abroad, 
by the Nishina Memorial Foundation,
and by NSF Grant PHY94-07194.

\begin{appendix}
\setcounter{equation}{0}
\section{A symmetry among the orthonormal polynomials}\l{appA}

In this appendix we prove a lemma relating polynomials orthonormal with
respect to measures $d\la w(\la)$ and $d\la\la^2 w(\la)$. It follows from
a more general statement in the book of Szeg\"{o} (\cite{SZE}, theorem 3.1.4).
We have referred to this relation as the $\alpha$-symmetry in the main text.

\noi
{\sc Lemma}: Let $P_n^{(\alpha)}(\la)$ denote the polynomials orthonormal 
w.r.t. the measure 
\beq
d\la w^{(\alpha)}(\la)\! \equiv\! d\la |\la|^{2\alpha} 
\exp (-NV(\la)) ~~~~~~,~\alpha>-\frac{1}{2} ~,~~~~~\la \in \mbox{\bf R} ~,
\eeq
where $V(\la^2)$ is an even polynomial 
with positive highest coefficient. The following identity then holds:
\beq
\frac{P_{2n+1}^{(\alpha)}(\la)}{\la} ~=~ P_{2n}^{(\alpha+1)}(\la) ~.
\eeq
{\sc Proof}: Since $\tilde{P}_{2n}^{(\alpha+1)}(\la) \equiv 
\frac{P_{2n+1}^{(\alpha)}(\la)}{\la}$ is an even polynomial of 
degree $2n$ it is orthonormal to all odd polynomials w.r.t. 
$d\la w^{(\alpha+1)}(\la)$. Because of 
\beqn
\int_{-\infty}^{\infty}\! d\lambda~ w^{(\alpha+1)}(\lambda) 
\tilde{P}^{(\alpha+1)}_{2m}(\lambda)\tilde{P}^{(\alpha+1)}_{2n}(\lambda) &=&
\int_{-\infty}^{\infty}\! d\lambda~ w^{(\alpha+1)}(\lambda) 
\frac{P^{(\alpha)}_{2m+1}(\lambda)P^{(\alpha)}_{2n+1}(\lambda)}{\la^2} 
\nonumber\\
&=& \int_{-\infty}^{\infty}\! d\lambda~ w^{(\alpha)}(\lambda) 
P^{(\alpha)}_{2m+1}(\lambda)P^{(\alpha)}_{2n+1}(\lambda) \nonumber\\
&=& \delta_{mn} 
\eeqn
one has
\beq
\tilde{P}_{2n}^{(\alpha+1)}(\la) ~=~ P_{2n}^{(\alpha+1)}(\la) ~.
\eeq

\noi
Inserting the above result into the definition (\r{defpsi}) 
of the wave functions $\psi_n(\la)$
immediately leads to the $\alpha$-symmetry as stated in eq. (\r{asym}).

\setcounter{equation}{0}
\section{The $m\geq 2$ multicritical case}\l{appB}

In this appendix we give the explicit result for the asymptotic form 
of the differential equation (\r{diff}) for $m=2$, and conjecture the form
for general $m$. The $m=2$ minimal multicritical potential reads 
\beq
V_2(\la) ~=~ \frac{1}{2}g_2\la^2 + \frac{1}{4}g_4\la^4 + \frac{1}{6}g_6\la^6
~,~~~~~~g_2 ~=~ -4 ~, ~~~~ g_4 ~=~ -16 ~, ~~~~ g_6 ~=~ 32 ~,
\eeq
with
\beq
\rho_2(\la) ~=~ \frac{1}{2\pi}g_6\la^4\sqrt{1-\la^2} ~.
\eeq
Performing the calculations exactly along the same lines as in subsection
\r{m=1expl} we get the following exact expressions
\beqn
A_n(\la) &=& N c_n\left( g_2 + g_4(\cnp + \cn) + g_6(\cnp(c_{n+2}^2+\cnp+\cn)+
                         \cn(\cnp+\cn+\cnm)) \right) \nonumber\\
          &&+ Nc_n\left( g_4 + g_6 (\cnp+\cn) \right)\la^2 
            ~+~ Nc_n g_6\la^4 \nonumber\\
B_n(\la) &=& N c_n^2\left( g_4 + g_6(\cnp+\cn+\cnm) \right)\la
             ~+~ N\cn g_6 \la ^3 ~+~ (1-(-1)^n)\frac{\alpha}{\la}
\eeqn
Inserting them into the exact from for $G_n(\la)$ eq. (\r{KFsGF})
and rescaling variables according to 
\beq
x ~=~ N^{\frac{1}{5}}\la ~,
\eeq
we expect that the following non-trivial contributions to be of order 1:
\beqn
u_{\pm} &\equiv& N^{\frac{4}{5}}\left( g_2 + g_4(c^2_{N\pm 1}+c^2_N) +
                 g_6(c^2_{N\pm 1}(c^2_{N\pm 2}+ c^2_{N\pm 1} + c^2_N)+
                     c^2_N(c^2_{N+1}+ c^2_N + c^2_{N-1})) \right) \nonumber\\
v ~~&\equiv& N^{\frac{3}{5}}\left( g_2 + 2g_4c^2_N +
                 2g_6c^2_N(c^2_{N+1}+ c^2_N + c^2_{N-1}) \right) \nonumber\\
q_{\pm} &\equiv& N^{\frac{2}{5}}\left( g_4 + g_6(c^2_{N\pm 1}+c^2_N) \right)
\nonumber\\
p ~~ &\equiv& N^{\frac{1}{5}}\left( g_4 + 2g_6c^2_N \right) \label{uvm=2}
\eeqn
This can be checked explicitly from the string equation for $V_2(\la)$ when 
inserting the ansatz (\r{moorexp}) 
for $m=2$. We do not display the differential
equations for the scaling functions $f_2(z),~g_2(z),~h_2(z)$ etc. here.
Finally the asymptotic expression for the
differential equation of the wave functions reads in the microscopic large-$N$ 
limit
\beq
\psi_N''(x) - N^{-1/5}F_N(x)\psi_N'(x) + N^{-2/5}G_N(x) \psi_N(x) ~=~ 0 ~,
\eeq
with
\beqn
N^{-1/5}F_N(x) &=& \frac{2q_+x ~+~ 4g_6x^3}{u_+ + q_+x^2 + g_6x^4}
\nonumber\\
N^{-2/5}G_N(x) &=& c_N^2u_+u_- + ((-1)^N\alpha -\frac{3}{2})v +
\left( c_N^2(u_+q_- + u_-q_+) -\frac{v^2}{4} + 
       ((-1)^N\alpha -\frac{1}{2})p\right)x^2  \nonumber\\
&+&\frac{(2u_+ + q_+x^2)(v+px^2) + (-1)^N2\alpha(q_+ + 2g_6x^2)}{u_+ + q_+x^2 +
         g_6x^4} + c_N^2g_6^2x^8 + \frac{(-1)^N\alpha -\alpha^2}{x^2} 
\l{FG2}
\eeqn
where only the dominant contributions are displayed. Moreover we have used
that the quartic and sextic terms in $G_N(x)$ are subdominant due to the 
following identities, which are valid up to subleading terms:
\beqn
0 &=& 2c_N^2 g_6(u_+ + u_-) + 2c_N^2q_+q_- - pv \nonumber\\
0 &=& 4c_N^2 g_6(q_+ + q_-) -p^2 ~.
\eeqn
The former follows directly from the string equation (\r{string})
whereas the latter can be seen to be valid when expressing it in terms of $v$.

\noi
For the general case of $m$th multicriticality 
we obtain the following form:
\beq
\psi_N''(x) - N^{-\frac{1}{2m+1}}F_N(x)\psi_N'(x) 
+ N^{-\frac{2}{2m+1}}G_N(x) \psi_N(x) ~=~ 0 ~.
\label{diffex}
\eeq
The $F_N$-term does not simplify:
\beq
N^{-\frac{1}{2m+1}}F_N(x) ~=~ 
\frac{N^{-\frac{2}{2m+1}}A_N'(x)}{N^{-\frac{1}{2m+1}}A_N(x)} ~,
\eeq
where $N^{-\frac{1}{2m+1}}A_N(x)$ is a polynomial in $x=N^{\frac{1}{2m+1}}\la$
of the order $2m$. The $G_N$-term which we conjecture to be of the following
form will simplify considerably:
\beq
N^{-\frac{2}{2m+1}}G_N(x) ~=~ Q(x) + \frac{R(x)}{N^{-\frac{1}{2m+1}}A_N(x)} 
  + c^2_Ng_{2m+2}x^{4m} + \frac{(-1)^N\alpha -\alpha^2}{x^2} ~.
\eeq
Here $Q(x)$ and $R(x)$ are both polynomials of order $2m-2$ in $x$. They
depend on certain universal combinations of coupling constants and recursion 
coefficients, as it has been explicitly shown in the two examples for 
$m=1$ and $m=2$ in eq. (\r{Gasympt}) and (\r{FG2}). The latter can be 
brought into the conjectured form when dividing out the quartic term in the 
denominator.

\end{appendix}

\end{document}